\documentclass[journal=cmatex,manuscript=article]{achemso}

\usepackage{chemformula} % Formula subscripts using \ch{}
\usepackage[T1]{fontenc}

\author{Ezra Alexander}
\date{\today}
\email{ezraa@mit.edu}
\author{Alexandra Alexiu}
\author{Matthias Kick}
\author{Troy Van Voorhis}
\affiliation{Department of Chemistry, Massachusetts Institute of Technology, Cambridge, Massachusetts 02139, USA}
\keywords{}

\title{Structural Hole Traps in III-V Quantum Dots}

\begin{document}

%%%%%%%%%%%%%%%%%%%%%%%%%%%%%%%%%%%%%%%%%%%%%%%%%%%%%%%%%%%%%%%%%%%%%
%% The "tocentry" environment can be used to create an entry for the
%% graphical table of contents. It is given here as some journals
%% require that it is printed as part of the abstract page. It will
%% be automatically moved as appropriate.
%%%%%%%%%%%%%%%%%%%%%%%%%%%%%%%%%%%%%%%%%%%%%%%%%%%%%%%%%%%%%%%%%%%%%
\begin{tocentry}

\includegraphics[width=\textwidth]{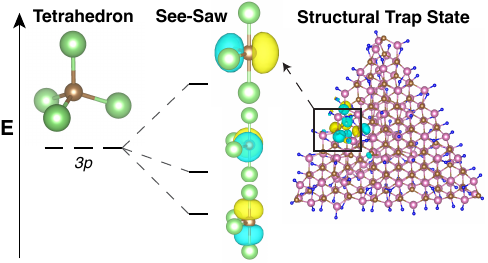}
For Table of Contents Only

\end{tocentry}

%%%%%%%%%%%%%%%%%%%%%%%%%%%%%%%%%%%%%%%%%%%%%%%%%%%%%%%%%%%%%%%%%%%%%
%% The abstract environment will automatically gobble the contents
%% if an abstract is not used by the target journal.
%%%%%%%%%%%%%%%%%%%%%%%%%%%%%%%%%%%%%%%%%%%%%%%%%%%%%%%%%%%%%%%%%%%%%
\begin{abstract}
Non-toxic III-V quantum dots (QDs) are plagued with a higher density of performance-limiting trap states than II-VI and IV-VI QDs. Such trap states are generally understood to arise from under-coordinated atoms on the QD surface. Here, we present computational evidence for, and an exploration of, trap states in InP and GaP QDs that arise from fully-coordinated atoms with distorted geometries, denoted here as structural traps. In particular, we focus on the properties of anion-centered hole traps, which we show to be relatively insensitive to the choice of the (typically cation-coordinating) ligand. Through interpolation of trap center cutouts, we arrive at a simple molecular orbital (MO) argument for the existence of structural traps, finding two main modalities: bond stretches and angular distortion to a see-saw-like geometry. These structural trap states will be important for understanding the low performance of III-V QDs, as even core-shell passivation may not remove these defects unless they can rigidify the structure. Moreover, they may lead to interesting dynamical properties as distorted structures could form transiently.
\end{abstract}

%%%%%%%%%%%%%%%%%%%%%%%%%%%%%%%%%%%%%%%%%%%%%%%%%%%%%%%%%%%%%%%%%%%%%
%% Start the main part of the manuscript here.
%%%%%%%%%%%%%%%%%%%%%%%%%%%%%%%%%%%%%%%%%%%%%%%%%%%%%%%%%%%%%%%%%%%%%
\section{Introduction}

In the past decade, colloidal semiconductor nanocrystals, better known as quantum dots (QDs), have become leading optoelectronic materials in fields ranging from quantum information\cite{vajner_quantum_2022,zajac_resonantly_2018} to photocatalysis\cite{sun_recent_2023,chen_mechanistic_2023} to biomedical imaging\cite{gidwani_quantum_2021,park_metallic_2022} to display technologies\cite{cui_advances_2023,wang_development_2023}. In particular, indium phosphide (InP) QDs have established themselves as the leading non-toxic\cite{campalani_towards_2025} alternative to Cd-based QDs in LED devices\cite{wang_development_2023,stam_near-unity_2024} due to their high brightness, stability and tunability\cite{yuan_near-infrared-absorbing_2025,won_highly_2019,li_stoichiometry-controlled_2019}. However, it has long been known\cite{guzelian_synthesis_1996,fu_inp_1997} that InP QDs natively have an extremely low photoluminescence quantum yield (PLQY)\cite{cao_layer-by-layer_2018}, with broad emission below the expected absorption onset\cite{hughes_effects_2019}. These effects are typically attributed to trap states:\cite{hughes_effects_2019,sung_increasing_2021} localized electronic states that manifest as mid-gap energy levels and serve as "stepping-stones" for the non-radiative and/or red-shifted recombination of charge carriers \cite{richter_fast_2019,janke_origin_2018}. While cutting-edge surface passivation techniques have paved the way to near-unity PLQY InP QDs\cite{stam_near-unity_2024,jo_highly_2021,li_stoichiometry-controlled_2019,lee_effectual_2022}, the fundamental atomistic nature of the trap states in these materials remains uncertain.

Trap states in III-V quantum dots\cite{kirkwood_finding_2018,hughes_effects_2019,stein_luminescent_2016,cho_optical_2018,kim_trap_2018,dumbgen_shape_2021,janke_origin_2018,enright_role_2022,schiettecatte_enhanced_2024,gwak_highly_2024,alexander_understanding_2024,zhu_boosting_2023,stam_guilty_2023,fu_inp_1997,ubbink_water-free_2022}, as well as in II-VI QDs\cite{houtepen_origin_2017,kirkwood_finding_2018,goldzak_colloidal_2021,mcisaac_it_2023,bhati_how_2023,elward_effect_2013,kilina_surface_2012}, IV-VI QDs \cite{xia_facet_2020,bender_surface_2018,zherebetskyy_tolerance_2015,voros_hydrogen_2017}, and lead halide perovskite nanocrystals \cite{nenon_design_2018,du_fosse_limits_2022}, are generally understood to arise from various under-coordinated surface atoms. The formation of trap states on under-coordinated surface atoms in InP has repeatedly been demonstrated with atomistic electronic structure simulations \cite{alexander_understanding_2024,cho_optical_2018,dumbgen_shape_2021,janke_origin_2018,stam_guilty_2023,ubbink_water-free_2022}, experimentally justified through correlation between trap density and ligand coverage \cite{kirkwood_finding_2018,hughes_effects_2019,schiettecatte_enhanced_2024} as well as with X-ray spectroscopies \cite{stein_luminescent_2016,kim_trap_2018,gwak_highly_2024}, and explained through the intuitive "dangling-bond" model \cite{fu_inp_1997}. However, there has historically been significant disagreement as to the relative importance of under-coordinated indium and phosphorus, as well as between the two-coordinate (-2c) and three-coordinate (-3c) species. Previously, we reported computational work exploring a library of 160 InP and GaP model QDs with different shapes, facetings, and surface defects using density functional theory (DFT) alongside novel orbital localization techniques \cite{alexander_understanding_2024}. In addition to characterizing the depth, relative frequency, and geometry dependence of the three-coordinate trap states in this library, this work also identified a significant proportion of trap states that arise, not from under-coordinated species, but instead from distorted but fully-coordinated atoms near the QD surface. These trap states, termed "structural traps," are generally more shallow than their dangling-bond counterparts, but further analysis is necessary to understand their origins and prevalence.

This core idea, that geometric distortions can infringe upon the band gap of semiconductors, is not new. Impacts to optical properties upon delocalized structural distortion have been computationally demonstrated in bulk semiconductors \cite{mishra_electronic_2016,mishra_effect_2012}, as well as experimentally realized in strained nanocrystals \cite{baimuratov_optical_2017,oksenberg_large_2020} and quantum dots \cite{cao_crystal_2015,bertolotti_crystal_2016,isarov_rashba_2017,dufour_halide_2019,moscheni_size-dependent_2018}. Localized structural distortions in real bulk semiconductors \cite{dolabella_lattice_2022,bozin_entropically_2010} and quantum dots \cite{guzelturk_dynamic_2021,kim_critical_2020} have also been detected experimentally. Such distortions have been shown to be ligand-dependent in many systems \cite{bertolotti_crystal_2016,kim_critical_2020,jana_ligand-induced_2017,dufour_halide_2019}. However, to the best of our knowledge, this is the first atomistic computational demonstration that trap states can be formed in quantum dots upon localized geometric distortions. It should be noted here that these distortions were not intentionally engineered, nor did they form in response to any sort of external perturbation; they are simply present in the optimized ground-state geometry of our models.

Here, we extend our previous investigation to characterize the prevalence and mechanistic origins of the different types of structural trap states in InP and GaP QDs. To further generalize our results, we more than double our original library of DFT-computed QD models to 360 structures. After discussing representative examples of indium- and phosphorus-localized structural trap states, we delve into their relative prevalence in different models. As the vast majority of indium and gallium structural trap centers in our models are bound directly to ligands, we expect their prevalence and properties to be somewhat dependent on our choice of passivation strategy. Therefore, we turn our focus to phosphorus-localized structural trap states, as we expect them to be more ligand-agnostic. We then present a simple MO picture of the mechanistic origins of structural trap states, validating it through interpolations of cutouts of each structural trap center in our database. These interpolations are, in turn, justified by computing the agreement between the interpolated MOs and the real QD trap states, showing that over 80\% of phosphorus-based structural traps are well described by a single center. Two main distortion modalities serve to explain these trap states: bond stretching and angular distortion from tetrahedral geometries toward a see-saw-like final geometry. These groupings are further supported by unsupervised machine learning; specifically, feature vectors are constructed and passed through dimensionality reduction\cite{maaten_visualizing_2008} followed by hierarchical clustering \cite{campello_hierarchical_2015}. We further find that geometry-only features are insufficient to accurately separate structural trap centers from non-trapping phosphorus, as a significant proportion of the latter have comparable distortions to the former. However, by introducing only simple features directly obtainable from DFT calculations, we are able to train a gradient-boosted trees (GBT) classifier \cite{friedman_greedy_2001} that can predict structural trap centers with an f1-score of over 90\%. This classifier validates our assigned labels and sheds further light on why some distorted, fully-coordinated atoms do not form structural trap states. This work opens the door for future experiments and simulations to explore the extent to which structural trap states contribute to the low performance of III-V QDs within different passivation schemes and the potentially dynamical mechanistic role of these trap states. 

\section{Computational Methodology}

The construction of our QD models has been discussed in detail elsewhere\cite{alexander_understanding_2024}. To summarize, we focus on core-only InP and GaP QDs, with 6 base morphologies for each material passivated with X-type \ch{F-} ligands \cite{hughes_effects_2019,kim_trap_2018}. Base morphologies are carved from the bulk crystal and passivated using a well-established procedure \cite{goldzak_colloidal_2021}. These QDs can generally be separated into two base geometries (three with a cuboctahedral geometry and three with a tetrahedral geometry) and two size ranges (two that are roughly 1.7-1.8 nm in diameter and four that are roughly 2.0-2.5 nm in diameter). Their surfaces are then diversified by the systematic creation of surface \ch{F-}, \ch{P^3-}, \ch{InF_x}, and \ch{InP} vacancies and subsequently relaxed to obtain the final 360 QD models. To overcome the limitations in available structures imposed by strict charge neutrality, we employ slight positive charges in some structures to maintain charge-orbital balance \cite{voznyy_charge-orbital_2012,zherebetskyy_tolerance_2015,dumbgen_shape_2021}.

The optimized geometries and band structures of our QD models have been determined at the DFT level in the CP2K and Q-Chem software packages\cite{kuhne_cp2k_2020,shao_advances_2015}, respectively. While geometries are optimized with the PBE functional, we use the more accurate PBE0 hybrid functional when computing the electronic structure as the incorporation of exact exchange has been shown to greatly improve predicted QD band gaps \cite{kurth_molecular_1999,azpiroz_benchmark_2014}. Karlsruhe effective core potentials are used to reduce computational cost and incorporate scalar relativistic effects \cite{weigend_balanced_2005}. While charge trapping is an inherently excited-state phenomenon \cite{goldzak_colloidal_2021}, the identification of localized mid-gap single-particle states in ground-state calculations has repeatedly been demonstrated to serve as an acceptable cost-efficient alternative \cite{houtepen_origin_2017,kim_trap_2018,cho_optical_2018,janke_origin_2018}. However, DFT's eigenstates do not have a well-defined localization \cite{boys_construction_1960,lehtola_unitary_2013}, leaving previous computational studies unable to clearly identify shallow trap states that have become unrepresentatively delocalized  in dense regions of the eigenspectrum \cite{truhlar_are_2012}. Our previous work overcomes this limitation by incorporating Pipek-Mizey orbital localization\cite{pipek_fast_1989} to algorithmically determine the valence band maximum (VBM) and conduction band minimum (CBM), and thus a complete set of trap states. This procedure is discussed in more detail in the original work \cite{alexander_understanding_2024}.

The interpolations presented in this work are accomplished by first determining the center of each structural trap state as the phosphorus with the highest in-state L{\"o}wdin population \cite{lowdin_nonorthogonality_1950}. Its geometry is then cut out and cations are replaced with lithium, with bond angles preserved and bond lengths adjusted accordingly, resulting in \ch{PLi4+} clusters. Interpolations are then performed from an ideal tetrahedral geometry to the distorted QD geometry using a 100-frame geodesic interpolation to ensure they remain within feasible space \cite{zhu_geodesic_2019}. The electronic structure at each point in the interpolation trajectory is then computed at the same PBE0/Def2-SVP level of theory as our QD models. Within this framework, the three highest occupied molecular orbitals (HOMOs) represent the phosphorus 3p orbitals, and the HOMO energy, relative to that of the HOMO-1 and HOMO-2, is representative of the depth of the associated trap state \cite{houtepen_origin_2017,alexander_understanding_2024}. Orbital-atom overlap is quantitatively measured by integrating the MO density within the covalent radii of each bound atom, and agreement between interpolated and QD MOs is quantitatively measured as the cosine similarity between the orientations of the p-like orbitals relative to their internal reference frames.

All ML analyses have been performed using scikit-learn \cite{pedregosa_scikit-learn_2011}. Full details of the featurization and hyper-parameters used in this study can be found in the Supporting Information (SI III.I and III.II). For the clustering, feature vectors for all P-4c structural trap centers were tested with both t-distributed Stochastic Neighbor Embedding (t-SNE)\cite{maaten_visualizing_2008} and the Uniform Manifold Approximation and Projection \cite{mcinnes_umap_2020}. Hierarchical Density-Based Spatial Clustering of Applications with Noise (HDBSCAN)\cite{campello_hierarchical_2015} was then used to determine clusters within the reduced space. For the classification, all P-4c were fed into a binary classification task, with the following architectures tested in both the geometry-only and DFT-included cases: support vector classifier\cite{platt1999probabilistic}, random forest \cite{breiman_random_2001}, GBT \cite{friedman_greedy_2001}, feed-forward neural networks with 1,3, and 5 hidden layers \cite{hinton_connectionist_1990}, and t-SNE- and UMAP-based approaches. As the two classes in this problem are heavily imbalanced, minority oversampling techniques such as SMOTE\cite{chawla_smote_2002} and ADASYN\cite{he_adasyn_2008} were tried, as was bagging\cite{breiman_pasting_1999}, but none of these techniques were found to significantly improve performance.

\section{Results and Discussion}

\subsection{Overview of Structural Trap States}

\begin{figure}[h]
\centering
\includegraphics[width=\textwidth]{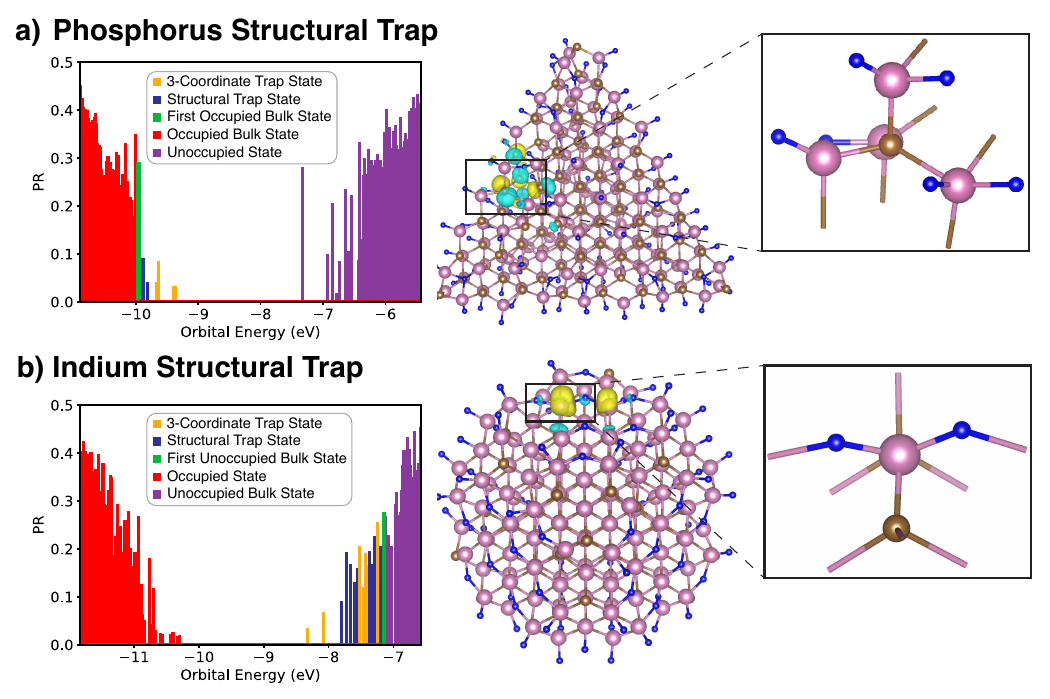}
\caption{Examples of phosphorus- (a) and indium- (b) localized structural trap states. Each displays a labeled participation ratio (PR) (left), plotted molecular orbital of the deepest structural trap state at an isosurface level of 0.03 (center), and a zoomed in cutout of that trap center (right). Participation ratio plots only label trap and first bulk states in the relevant band for the corresponding central element. Pink spheres represent indium, brown spheres represent phosphorus, and blue spheres represent fluorine.}
\end{figure}

Structural trap states are present in 301 of the 360 model QDs studied here. In this work, we define a structural trap state as any trap state, i.e a well-localized state between the identified VBM and CBM, where the largest individual in-state L{\"o}wdin population is associated with an atom with a coordination number  of four or greater. That atom is then identified as that trap state's center (SI I.I). Note that this will in-general underestimate the "true" number of structural trap states in our QD models, as the mixing between states in dense energy windows will often result in a single under-coordinated atom being the dominant contribution to multiple adjacent states. Also note that, in principle, this definition allows homo-atomic dimers and atoms with coordination numbers of five or more to be considered structural trap centers. While such defects do occasionally occur within our dataset, both are quite rare and usually do not form structural trap states even when they do occur (SI I.II). Nevertheless, they are excluded from future discussion.

A representative example of a P-4c structural trap can be found in Figure 1a. This particular QD, a roughly 2.5 nm InP tetrahedron with an induced \ch{InP} vacancy, has 6 hole traps: four localized on P-3c and two P-4c structural traps. These structural traps have depths of 0.20 and 0.13 eV and are localized in the outermost monolayer of the QD, likely as an indirect result of the relaxation associated with the nearby \ch{InP} vacancy. More specifically, the P-4c center of the trap state (Figure 1a, right) has had one of its bonds stretched to roughly 1.15 times its original bond length. As is the case in the example, P-4c structural trap states are generally less deep across our dataset (average 0.19 eV) than P-3c trap states (average 0.59 eV). However, the deepest P-4c structural trap states have depths approaching 1 eV, and 39 of our 360 QD models have a P-4c structural trap as their deepest hole trap. Moreover, certain QDs without induced vacancies display P-4c structural trap states.

A representative example of an In-4c structural trap can be found in Figure 1b. This particular QD, a roughly 2.3 nm InP truncated cuboctahedron with an induced \ch{F-} vacancy, has 17 electron traps: 8 localized on In-3c and 9 In-4c structural traps. These structural traps appear to be somewhat independent of the \ch{F-} vacancy induced in this structure, occurring regularly along the (1 0 0) facets of our cuboctahedral QDs on In-4c with stretched F-In-F bond angles. The In-4c displayed on the right of Figure 1b, for example, has a F-In-F bond angle of 135 degrees. In fact, only around 18\% of the In-4c structural traps in our database occur on sub-surface In not bound to fluorine. While many In-3c are non-trapping \cite{alexander_understanding_2024}, trapping In-3c (0.64 eV) are on average only moderately deeper than In-4c structural traps (0.48 eV).

Important structural differences between InP and GaP QDs mean that structural traps in GaP tend to look somewhat different than those in InP. In particular, InP QDs are far more labile than GaP QDs \cite{alexander_understanding_2024}, with large displacements of surface atoms upon structural relaxation. This results in far more structural traps in InP QDs than GaP QDs on average, though the smaller size of Ga atoms makes bond stretches somewhat more prevalent in GaP QDs than InP QDs. Outside of these geometric differences, P-4c structural traps in GaP seem to behave analogously to those in InP.  Examples and further discussion of structural trap states in GaP QDs can be found in the Supporting Information (SI I.III).

One reasonable challenge to these results is the reliance of our models on \ch{F-} ligands, where the small halide is more likely to induce strain on the QD surface when bridging two cations. Another reasonable challenge is the presence of many under-coordinated trap states in our QDs, which may be miscategorized as structural trap states or indirectly cause such states in some other way. To address both of these challenges, we present two 2.7 nm cuboctahedral QDs, one InP and one GaP, that are perfectly passivated with \ch{Cl-} anions (SI I.III). These QDs are charge-neutral and highly symmetric, containing no atoms with coordination numbers below four and no strong internal electric fields. Nevertheless, both QDs display multiple clearly-localized shallow hole traps, with the deepest in InP having a depth of 0.25 eV. These trap states are associated with surface-monolayer P-4c, with those in InP having relatively slight distortions, suggesting the difficulty of fully eliminating P-4c structural traps. Nevertheless, this \ch{Cl-} passivated QD does not seem to have any In or Ga structural traps. The observation that the anionic structural traps persist across different passivating ligands while cationic structural traps are more ligand-dependent motivates us to focus on the anionic traps below.

\subsection{Prevalence of Structural Traps}

\begin{figure}[!]
\centering
\includegraphics[width=0.5\textwidth]{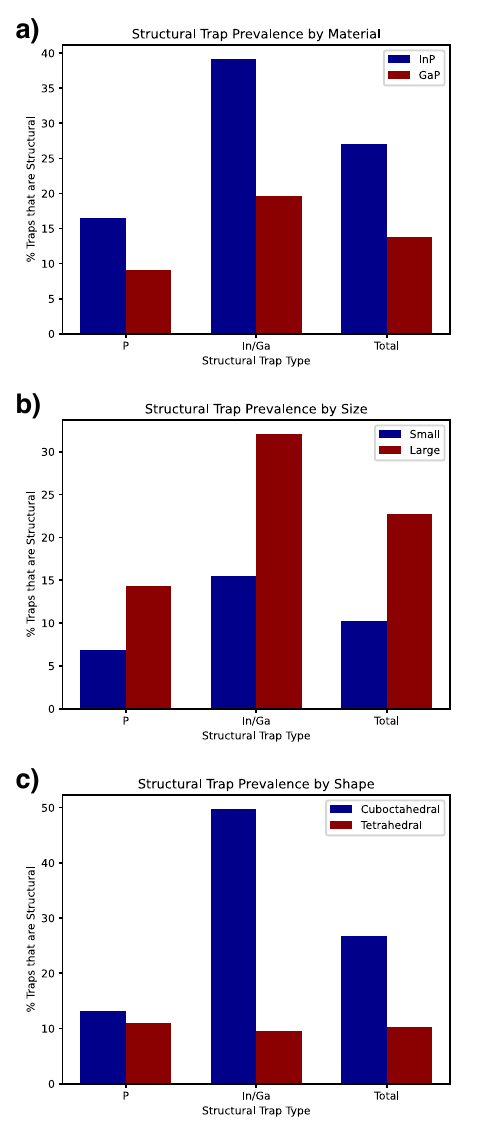}
\caption{Percentage of total trap states that are structural traps by (a) material, (b) QD size, and (c) QD shape. In each figure, the leftmost set of columns represents phosphorus-localized hole traps, the central column represents cation-localized electron traps, and the rightmost figure represents all trap states.}
\end{figure}

While structural trap states are generally less frequent than under-coordinated trap states across our dataset, certain factors found to influence these proportions are summarized in Figure 2. In general, we find that cation-centered structural electron traps make up a larger percentage of total electron traps (29\%) than the phosphorus-localized variety makes up of total hole traps (12\%). This effect is two-fold, in that there are both more cation-centered structural traps (1,084) than phosphorus-centered structural traps (578) and fewer total electron traps (2,782) than total hole traps (4,051). In terms of percentage of total atoms, 2.1\% of P-4c, 2.4\% of Ga-4c, and 3.6\% of In-4c in our dataset act as structural trap centers.

As discussed above, both phosphorus- and cation-centered structural traps are more prevalent in InP than GaP (Figure 2a) by a factor of roughly 2. We have also already discussed the increased prevalence of electron traps on the (1 0 0) facets of our cuboctahedral QD models (Figure 2c). Surprisingly, we also find that both electron and hole structural traps make up a larger percentage of the trap states in our larger QDs (2 - 2.5 nm in diameter) than they do in our smaller QDs (1.7-1.8 nm in diameter). This is also true when considering the percentage of total 4c atoms which are structural trap centers in our small (2.0\%) and large (2.7\%) QD models, despite the fact that the surface monolayer becomes a smaller proportion of the total QD. While this latter effect will likely fall off above a certain size range, the proportion of total trap states which are structural in these larger QDs implies that such structural traps should remain relevant even in experimental QD size ranges.

While In and Ga structural trap centers play a large role in our QD models, we are unable to fully separate these results from the \ch{F-} ligands that passivate our QD surfaces. However, P-4c structural traps remain a significant percentage of the total hole traps in our QDs and we have shown them to be present even in \ch{Cl-} passivated QDs. As such, we will focus only on phosphorus-localized structural hole traps going forward.

\subsection{Interpolation of P-4c Structural Trap Centers}

\begin{figure}[!]
\centering
\includegraphics[width=\textwidth]{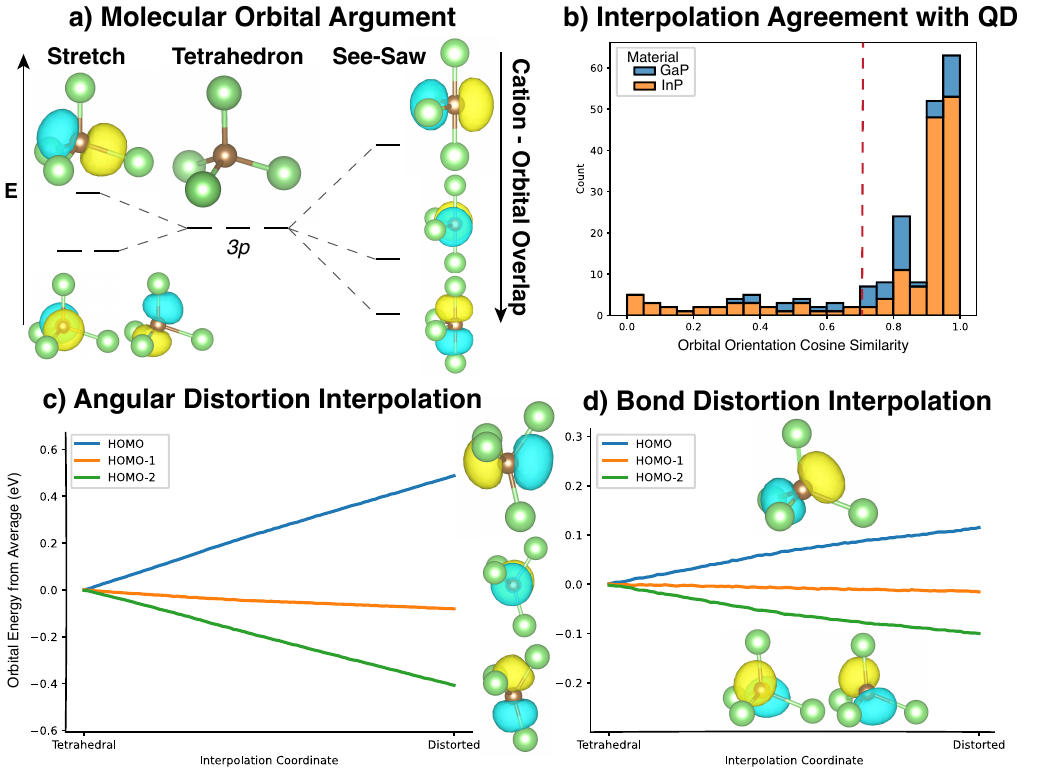}
\caption{Interpolations of \ch{PLi4+} cutouts. (a) Proposed Walsh diagram for the splitting of P 3p orbitals in \ch{PLi4+} upon bond stretch (left) and angular distortion to a see-saw geometry (right). Inset molecular orbitals are taken directly from idealized models. (b) Histogram of agreement between interpolated \ch{PLi4+} HOMOs and the relevant QD trap states, separated by material and measured by cosine similarity. A value of 0.707, representative of an angle of 45 degrees, is used as the cutoff point for acceptable agreement. (c) Interpolation of calculated orbital binding energies from a perfect tetrahedron to a see-saw-like geometry taken from a real QD, with a maximum angle of around 140 degrees. (d) Interpolation of calculated orbital binding energies from a perfect tetrahedron to a stretched bond geometry taken from a real QD, with a maximum bond stretch of around 1.15x. Brown spheres represent phosphorus and green spheres represent lithium. All molecular orbitals are plotted at an isosurface level of 0.06. }
\end{figure}

To understand the extent to which P-4c structural trap states can be explained by single-center geometric distortions, we have performed interpolations between an idealized tetrahedral geometry and the geometry of the center of each P-4c structural trap in our dataset (Figure 3). We find \ch{PLi4+} to serve as a functional model system for the geometry of the central phosphorus (SI II.I). As we interpolate from pristine to distorted geometries, we generally find that the HOMO, which corresponds to the trap state, increases in energy by 0.1 - 0.4 eV, with an average increase of 0.25 eV. This interpolated depth, which serves to raise the orbital into the band gap, tends to correlate with the actual depth of the associated trap state (SI II.II). 

Geometric distortions can be broadly separated into angular distortions and bond length distortions, though many P-4c with the latter also display some extent of the prior. Bond distortions (Figure 3d) are associated with moderate splittings in orbital binding energies, with more stretched bonds leading to larger splittings. As the bond is stretched, the HOMO increasingly occupies the space abandoned by the leaving cation, approaching the traditional "dangling bond" picture well understood for under-coordinated trap states. While cutouts from our QD dataset rarely maintain perfect symmetry, in an idealized stretching distortion (Figure 3a, left) the remaining two p orbitals remain degenerate and decrease in energy relative to the average. We find that the majority of P-4c angular distortions in our dataset seem to approach a see-saw-like geometry. While none achieve a full $180^\circ$ bond angle, they tend to include a single highly stretched bond angle ($130^\circ - 145^\circ$) as well as a roughly $120^\circ$ bond angle and several bond angles approaching $90^\circ$. This is further reflected by a characteristic orbital splitting in their interpolations (Figure 3c), wherein the HOMO increases in energy by 0.3-0.5 eV, the HOMO-1 decreases in energy slightly, and the HOMO-2 decreases in energy by a corresponding amount to the HOMO. This splitting and the associated orbital positions reflect those seen in an idealized see-saw geometry, wherein the HOMO bisects the $180^\circ$ bond angle (Figure 3a, right).

The governing principle behind both of these modalities appears to be the extent to which each P 3p orbital overlaps with the bonded cations. In each case, the HOMO associated with the trap state seems to occupy the pocket that minimizes this overlap, with lower overlaps associated with higher splittings. This effect can be quantitatively measured by computing the overlap of each P 3p orbital with the covalent radii of each bonded cation. Upon doing so, with few exceptions this proposed hierarchy of decreasing overlap with increasing energy is found to hold true for every interpolation studied here (SI II.III). This effect can be easily understood through the same arguments used in crystal field theory \cite{PhysRev.41.208}. Whereas in traditional crystal field theory electronic orbitals are \textit{destabilized} by overlap with anions bonded to a metal center, here they are \textit{stabilized} by overlap with cations bonded to an anionic center.

The extent to which the orbitals predicted by these interpolation-based arguments are reflected in the actual QD models can be quantitatively measured by determining a corresponding orientation vector for the central P 3p orbital in each and measuring their cosine similarity (Figure 3b). This metric will in general fail in the regimes where multiple states have become highly mixed, however, and such structural traps are excluded from this analysis. Of the remaining interpolations, 80.2\% agree closely with the orbital orientation in their respective QDs.  We consider the remaining structural traps to be poorly described by the geometric variations of a single center. One effect clearly missing from these interpolations that may explain the remaining 19.8\% is the internal electric field of the QD. We have explored schemes to incorporate these fields into our interpolations (SI II.IV), and while such effects can serve to explain the increased depth of certain otherwise shallow interpolations, they do not in general improve the relative orientation of the interpolated MOs.

\subsection{Clustering of P-4c Structural Trap Types}

\begin{figure}[h]
\centering
\includegraphics[width=\textwidth]{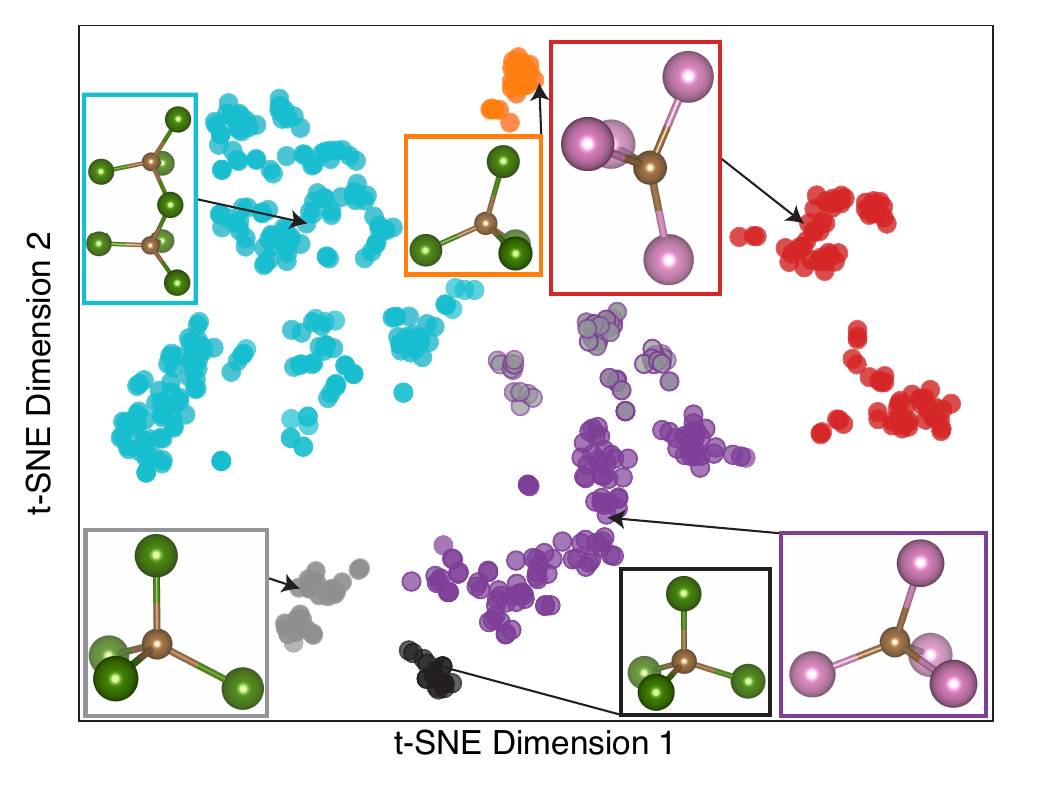}
\caption{T-distributed stochastic neighbor embedding of feature vectors for all P-4c structural trap centers, wherein high-dimensional data has been projected into two dimensions while approximately maintaining relative distances. Clusters obtained from HBSCAN clustering have been grouped and colored by human intuition to present more generalized results. Representative structures from each cluster are inset, with pink spheres representing indium, brown spheres representing phosphorus, and green spheres representing gallium.}
\end{figure}

Insights into the distribution and prevalence of different types of structural trap centers can be obtained using unsupervised machine learning. We consider 8 categories of features to describe structural trap centers in this study: bond lengths, bond angles, QD composition, electric field, MO energies, partial charges, projected density of states (pDOS), and PR (SI III.I). Of these, the first half are directly obtainable from a geometry file while the latter half are obtained from an electronic structure calculation. While we have considered more delocalized descriptions of the geometric environment of the trap centers, they are found to consistently decrease the performance of the following classification task, and are thus excluded here for consistency. While our labels of trap or bulk are obtained using the orbital localization procedure described previously, any features that can only be obtained from that procedure, such as the energetic position of the VBM and CBM, are excluded here. In the case of this unsupervised task, all features are used save for those shared by all structural traps in a given QD,  as they otherwise dominate clustering. 

Through these features, all P-4c structural trap data is visualized in Figure 4. PCA-initialized t-SNE is used to project these features into 2D space, and then HDBSCAN clustering is used to group and label clusters. The results of the HDBSCAN clustering can be found in the Supporting Information (SI III.III); in Figure 4 we combine certain clusters into larger groups when appropriate for ease of viewing. This results in seven distinct clusters: three  that represent see-saw like distortions, two that represent bond stretches, and two that represent difficult-to-characterize structural trap centers. In general, structures near the top-right of the plot have the highest degree of angular distortion, while structures near the bottom have the lowest. The red group represents deep and localized P-4c structural traps that are the closest to the proposed see-saw geometry, all in InP. The orange and purple groups correspondingly represent moderately distorted see-saw structures in GaP and InP, respectively, that remain well-described by the interpolations. Similarly, the smaller gray and gray-purple groups represent structures with bond stretches in the absence of and presence of additional angular distortion, respectively. While the gray group only contains structures in GaP, the gray-purple group contains a mix of InP and GaP QDs and is unsurprisingly associated with deeper trap states.

The two remaining groups, blue and black, represent two categories that fail to be captured in our interpolations. The black group consists of near-perfect tetrahedra in GaP that have been shifted into the band gap by extreme internal electric fields. While other structural trap states are doubtless deepened by electric fields, these effects tend to be somewhat lessened by the surface reconstructions in InP QDs. The blue group consists of low-moderate distortions in both InP and GaP that are delocalized over multiple centers. While the distortions in these centers are likely too moderate to cause structural trap states on their own, P 3p orbitals on adjacent sites tend to have an anti-bonding interaction that may serve to deepen the associated state. Limited interpolations of \ch{P2Li7+} cutouts have been performed that support this idea (SI II.V); however, a comprehensive exploration of this mechanism is beyond the scope of this work.

\subsection{Separating Trapping and Inert P-4c}

\begin{figure}[!]
\centering
\includegraphics[width=\textwidth]{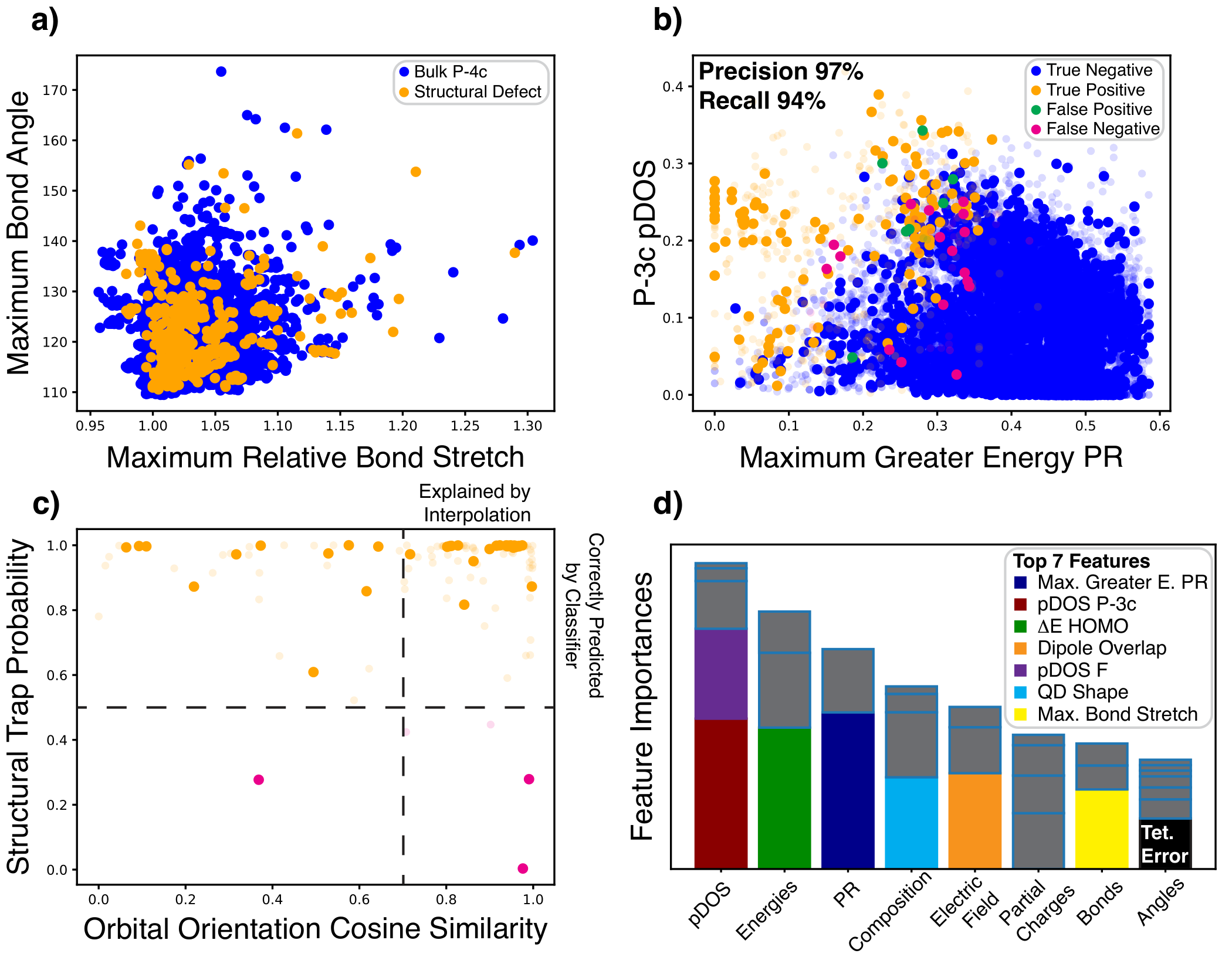}
\caption{Overview of full GBT classifier performance and operative features. (a) Problem overview: scatter plot of bulk and trapping P-4c in terms of bond and angular distortion metrics. (b) Scatter plot of GBT classifier performance in terms of the two highest impact features. A positive refers to a prediction that a P-4c is the center of a structural trap state. Inset precision and recall metrics refer to the unweighted class average. (c) Scatter plot comparison of classifier performance to interpolation orbital accuracy on appropriately localized structural trap states. Orange dots refer to correct predictions by the classifier while pink dots refer to incorrect predictions. Opaque points in (c) and (d) represent the test set while translucent points represent the test set. (d) Importances for all categories of features used in the GBT classifier, broken down by the importances of individual features. The top 7  individual features are represented in the order listed in the key. The 12th most important feature, the most important angle-based feature, is colored in black and labeled separately. Tet. Error stands for the summed absolute error of all bond angles from a tetrahedral geometry.
}
\end{figure}

Thus far, we have considered the differences between, and characteristics of, different types of structural trap states. An elimination of structural trap centers, however, necessitates some understanding of when they do not form. As shown in Figure 5a, there is unfortunately no clear divide between geometries that give rise to structural trap states and those that are inert. 73\% of structural trap centers have either a bond stretched by at least 1.05 or a maximum bond angle over 120, whereas this is true for only 38\% of bulk P-4c. However, there are many bulk P-4c that are highly distorted (the individual P-4c with the highest bond angle and bond stretch are both non-trapping), and there are many structural trap centers with near-negligible distortions. The creation of a tool that can separate bulk P-4c from structural trap centers accurately would not only be directly valuable, but would also validate our labels and shed light on the more subtle features which give rise to otherwise difficult to understand structural trap states.

Data-based determination of complex decision boundaries is one of the defining problems for machine learning. We formulate these tasks as a binary classification problem, complicated by a moderately sized (roughly 22,500 P-4c) and highly imbalanced dataset. A full discussion of architecture, feature, and hyperparameter selection for each task is presented in the Supporting Information (SI III.I and III.II). The weights and code for our best performing classifiers are freely available online \cite{ea_repo}. While we began by exploring the possibility of building a classifier that only utilizes features directly obtainable from a geometry file, the best classifier we are able to obtain this way displays a macro f1-score of only 0.74. This single-layer neural network is only accurate in 58\% of its predictions that a P-4c is a structural trap (precision), and only correctly identifies 45\% of structural traps in the test set (recall). It is always possible, however, that with more high-quality data a geometry-based separation between bulk and trapping P-4c could be found.

By introducing features obtained from DFT calculations, we are able to realize a gradient-boosted trees classifier that achieves a macro f1-score of 0.95. This classifier has significantly higher structural trap precision (94\%) than recall (87\%), as would be expected for an imbalanced learning task. As shown in Figure 5b, taking high-importance features from the classifier provides general visual separation between bulk and trapping P-4c inaccessible with only geometric features, and the classifier's mispredictions generally occur in these overlapping regions. These overlapping regions generally correspond with two-center structural trap states, the class which on which the interpolations also give the lowest agreement. To make the most direct comparison, we re-train and test the GBT classifier on the set of structural trap states for which the orbital orientation metric is well defined (Figure 5c) and achieve even higher performance (macro-f1 0.98). While the classifier is generally able to correctly label all but one of the multi-center structural traps for which the interpolations fail, there are a handful of particularly interesting P-4c for which the interpolations yield the correct orbital orientation but the classifier makes the wrong prediction. Ultimately, the high accuracy of the classifier in separating bulk and trapping P-4c makes it unlikely that a significant proportion of our structural traps have been mislabeled.

The method used to determine the importances of the 42 features used in the GBT classifier as well as full importance results are provided in the Supporting Information (SI III.IV). As expected, we find that the GBT classifier's performance depends most strongly on categories of features that must be obtained from a DFT calculation, such as the energy, pDOS, and localization of the associated MO (Figure 5d).  While the connection between the two highest-performing features and structural trap states may appear nebulous at first, both reflect less on the nature of a structural trap state and more on the nature of a general trap state: the state should be above the delocalized VBM in energy and will thus mix with other P-3c trap states. In fact, the top three features are all directly incorporated into our procedure for identifying the VBM. The next tier of features have more clear physical significance, as dipole overlap reflects the direction and magnitude of electric-field-induced state shifting and pDOS F reflects the surface-association of the state. The high importance of QD shape is surprising, as Figure 2c shows similar proportions of P-4c structural traps in both shape classes. Center bonds and angles are the two least important classes, likely due to the misleading overlaps shown in Figure 5a. While the maximum relative bond stretch comes in as the 7th most important feature, the most important angle-based feature, an aggregated angular error from a tetrahedral geometry, is only the 12th most important feature. While the difference in these importances is not massive, it is not surprising that the high-quality data available from a complex electronic structure calculation provides an easier identification of this complex phenomenon.

\section{Conclusions}

We have utilized a large and diverse library of DFT-computed QD electronic structures to explore the emergence of trap states localized on fully-coordinated atoms. These structural trap states are often clearly localized on distorted near-surface atoms, though they are generally less deep than three-coordinate trap states, and occur even in perfectly \ch{Cl-} passivated QDs. They make up a significant proportion of the total trap states in these structures, being more prevalent in our larger structures than our smaller ones. They occur more frequently on In and Ga than P, though this effect likely arises from \ch{F-} passivation. We have used interpolations to show that P-4c structural traps can be mechanistically explained through either bond stretches or angular distortions that decrease P 3p orbital overlap with bonded cations, destabilizing those MOs into the band gap. Many of these angular distortions approach a see-saw-like geometry with one dominant bond stretch, which has a specific p orbital splitting signature. However, not all P-4c structural traps are well explained by a single center's geometry. Both non-local electrostatic effects and consecutive slight distortions cannot be captured in a single interpolation, but emerge as classes of P-4c structural traps in t-SNE dimensionality reductions. They are, however, the minority, with the most common class being a spectrum of see-saw like angular distortions. We have further explored leveraging our dataset to train a classifier that can accurately separate bulk and trapping P-4c, but this is more difficult than it may seem. Geometric features overlap too heavily between the two classes, and so electronic structure features must be included to achieve good results. The resulting GBT classifier is able to accurately separate out even those structural traps which are poorly described by the interpolations by combining geometric and electrostatic information with a learned understanding of trap states.

These results reveal a crucial new modality of trap state that must be better understood if III-V trap states are to be fully controlled. Future directions include an extension of the InP(F) and GaP(F) materials studied here to a wider range of III-V materials and surface passivations. For example, it is possible that core-shell hetero-structures are efficient at eliminating not only under-coordinated sites but also structural traps, though structural traps could form when the lattice distortion is high. Similarly, surface oxidation and other impurities distort the structure of the III-V lattice, and the distinct effect of this distortion could potentially be understood in the context of structural trapping. Finally, the application of a full or approximate \cite{kick_super-resolution_2024} excited state electronic structure method to study the trap states in these QDs would give a better sense of the transferability of these ground state results.

%%%%%%%%%%%%%%%%%%%%%%%%%%%%%%%%%%%%%%%%%%%%%%%%%%%%%%%%%%%%%%%%%%%%%
%% The "Acknowledgement" section can be given in all manuscript
%% classes.  This should be given within the "acknowledgement"
%% environment, which will make the correct section or running title.
%%%%%%%%%%%%%%%%%%%%%%%%%%%%%%%%%%%%%%%%%%%%%%%%%%%%%%%%%%%%%%%%%%%%%
\begin{acknowledgement}

This work was supported by the U.S. National Science Foundation under grant number CHE-2154938. This work was also supported through an allocation on the SDSC Expanse through ACCESS Maximize project CHE200006.

\end{acknowledgement}

%%%%%%%%%%%%%%%%%%%%%%%%%%%%%%%%%%%%%%%%%%%%%%%%%%%%%%%%%%%%%%%%%%%%%
%% The same is true for Supporting Information, which should use the
%% suppinfo environment.
%%%%%%%%%%%%%%%%%%%%%%%%%%%%%%%%%%%%%%%%%%%%%%%%%%%%%%%%%%%%%%%%%%%%%
\begin{suppinfo}

The Supporting Information is available free of charge at XXX

\begin{itemize}
  \item Structural trap center identification, rare structural defects, structural trap states in GaP, perfectly passivated QDs, choice of \ch{PLi4+}, correlation with trap depth, quantitative atom-orbital overlap, interpolations in an external electric field, 2-center P-4c trap states, featurization, choice of architecture and hyper-parameters, HDBSCAN clustering, feature importances (PDF)
\end{itemize}

\end{suppinfo}

%%%%%%%%%%%%%%%%%%%%%%%%%%%%%%%%%%%%%%%%%%%%%%%%%%%%%%%%%%%%%%%%%%%%%
%% The appropriate \bibliography command should be placed here.
%% Notice that the class file automatically sets \bibliographystyle
%% and also names the section correctly.
%%%%%%%%%%%%%%%%%%%%%%%%%%%%%%%%%%%%%%%%%%%%%%%%%%%%%%%%%%%%%%%%%%%%%
\bibliography{bibliography}

\providecommand{\latin}[1]{#1}
\makeatletter
\providecommand{\doi}
  {\begingroup\let\do\@makeother\dospecials
  \catcode`\{=1 \catcode`\}=2 \doi@aux}
\providecommand{\doi@aux}[1]{\endgroup\texttt{#1}}
\makeatother
\providecommand*\mcitethebibliography{\thebibliography}
\csname @ifundefined\endcsname{endmcitethebibliography}  {\let\endmcitethebibliography\endthebibliography}{}
\begin{mcitethebibliography}{87}
\providecommand*\natexlab[1]{#1}
\providecommand*\mciteSetBstSublistMode[1]{}
\providecommand*\mciteSetBstMaxWidthForm[2]{}
\providecommand*\mciteBstWouldAddEndPuncttrue
  {\def\EndOfBibitem{\unskip.}}
\providecommand*\mciteBstWouldAddEndPunctfalse
  {\let\EndOfBibitem\relax}
\providecommand*\mciteSetBstMidEndSepPunct[3]{}
\providecommand*\mciteSetBstSublistLabelBeginEnd[3]{}
\providecommand*\EndOfBibitem{}
\mciteSetBstSublistMode{f}
\mciteSetBstMaxWidthForm{subitem}{(\alph{mcitesubitemcount})}
\mciteSetBstSublistLabelBeginEnd
  {\mcitemaxwidthsubitemform\space}
  {\relax}
  {\relax}

\bibitem[Vajner \latin{et~al.}(2022)Vajner, Rickert, Gao, Kaymazlar, and Heindel]{vajner_quantum_2022}
Vajner,~D.~A.; Rickert,~L.; Gao,~T.; Kaymazlar,~K.; Heindel,~T. Quantum {Communication} {Using} {Semiconductor} {Quantum} {Dots}. \emph{Advanced Quantum Technologies} \textbf{2022}, \emph{5}, 2100116\relax
\mciteBstWouldAddEndPuncttrue
\mciteSetBstMidEndSepPunct{\mcitedefaultmidpunct}
{\mcitedefaultendpunct}{\mcitedefaultseppunct}\relax
\EndOfBibitem
\bibitem[Zajac \latin{et~al.}(2018)Zajac, Sigillito, Russ, Borjans, Taylor, Burkard, and Petta]{zajac_resonantly_2018}
Zajac,~D.~M.; Sigillito,~A.~J.; Russ,~M.; Borjans,~F.; Taylor,~J.~M.; Burkard,~G.; Petta,~J.~R. Resonantly driven {CNOT} gate for electron spins. \emph{Science} \textbf{2018}, \emph{359}, 439--442\relax
\mciteBstWouldAddEndPuncttrue
\mciteSetBstMidEndSepPunct{\mcitedefaultmidpunct}
{\mcitedefaultendpunct}{\mcitedefaultseppunct}\relax
\EndOfBibitem
\bibitem[Sun \latin{et~al.}(2023)Sun, Xing, Li, and Zhou]{sun_recent_2023}
Sun,~P.; Xing,~Z.; Li,~Z.; Zhou,~W. Recent advances in quantum dots photocatalysts. \emph{Chemical Engineering Journal} \textbf{2023}, \emph{458}, 141399\relax
\mciteBstWouldAddEndPuncttrue
\mciteSetBstMidEndSepPunct{\mcitedefaultmidpunct}
{\mcitedefaultendpunct}{\mcitedefaultseppunct}\relax
\EndOfBibitem
\bibitem[Chen \latin{et~al.}(2023)Chen, Yu, Fan, Wu, and Zhou]{chen_mechanistic_2023}
Chen,~Y.; Yu,~S.; Fan,~X.-B.; Wu,~L.-Z.; Zhou,~Y. Mechanistic insights into the influence of surface ligands on quantum dots for photocatalysis. \emph{Journal of Materials Chemistry A} \textbf{2023}, \emph{11}, 8497--8514\relax
\mciteBstWouldAddEndPuncttrue
\mciteSetBstMidEndSepPunct{\mcitedefaultmidpunct}
{\mcitedefaultendpunct}{\mcitedefaultseppunct}\relax
\EndOfBibitem
\bibitem[Gidwani \latin{et~al.}(2021)Gidwani, Sahu, Shukla, Pandey, Joshi, Jain, and Vyas]{gidwani_quantum_2021}
Gidwani,~B.; Sahu,~V.; Shukla,~S.~S.; Pandey,~R.; Joshi,~V.; Jain,~V.~K.; Vyas,~A. Quantum dots: {Prospectives}, toxicity, advances and applications. \emph{Journal of Drug Delivery Science and Technology} \textbf{2021}, \emph{61}, 102308\relax
\mciteBstWouldAddEndPuncttrue
\mciteSetBstMidEndSepPunct{\mcitedefaultmidpunct}
{\mcitedefaultendpunct}{\mcitedefaultseppunct}\relax
\EndOfBibitem
\bibitem[Park \latin{et~al.}(2022)Park, Yang, Jung, Ko, Song, Hong, Kim, Lee, and Song]{park_metallic_2022}
Park,~K.~H.; Yang,~J.~Y.; Jung,~S.; Ko,~B.~M.; Song,~G.; Hong,~S.-J.; Kim,~N.~C.; Lee,~D.; Song,~S.~H. Metallic {Phase} {Transition} {Metal} {Dichalcogenide} {Quantum} {Dots} as {Promising} {Bio}-{Imaging} {Materials}. \emph{Nanomaterials} \textbf{2022}, \emph{12}, 1645\relax
\mciteBstWouldAddEndPuncttrue
\mciteSetBstMidEndSepPunct{\mcitedefaultmidpunct}
{\mcitedefaultendpunct}{\mcitedefaultseppunct}\relax
\EndOfBibitem
\bibitem[Cui \latin{et~al.}(2023)Cui, Yang, Qin, Wen, He, Mei, Zhang, Xing, Liang, and Guo]{cui_advances_2023}
Cui,~Z.; Yang,~D.; Qin,~S.; Wen,~Z.; He,~H.; Mei,~S.; Zhang,~W.; Xing,~G.; Liang,~C.; Guo,~R. Advances, {Challenges}, and {Perspectives} for {Heavy}-{Metal}-{Free} {Blue}-{Emitting} {Indium} {Phosphide} {Quantum} {Dot} {Light}-{Emitting} {Diodes}. \emph{Advanced Optical Materials} \textbf{2023}, \emph{11}, 2202036\relax
\mciteBstWouldAddEndPuncttrue
\mciteSetBstMidEndSepPunct{\mcitedefaultmidpunct}
{\mcitedefaultendpunct}{\mcitedefaultseppunct}\relax
\EndOfBibitem
\bibitem[Wang \latin{et~al.}(2023)Wang, Li, Chen, Lin, Niu, Yang, and Tang]{wang_development_2023}
Wang,~S.; Li,~Y.; Chen,~J.; Lin,~O.; Niu,~W.; Yang,~C.; Tang,~A. Development and challenges of indium phosphide-based quantum-dot light-emitting diodes. \emph{Journal of Photochemistry and Photobiology C: Photochemistry Reviews} \textbf{2023}, \emph{55}, 100588\relax
\mciteBstWouldAddEndPuncttrue
\mciteSetBstMidEndSepPunct{\mcitedefaultmidpunct}
{\mcitedefaultendpunct}{\mcitedefaultseppunct}\relax
\EndOfBibitem
\bibitem[Campalani and Monbaliu(2025)Campalani, and Monbaliu]{campalani_towards_2025}
Campalani,~C.; Monbaliu,~J.-C.~M. Towards sustainable quantum dots: {Regulatory} framework, toxicity and emerging strategies. \emph{Materials Science and Engineering: R: Reports} \textbf{2025}, \emph{163}, 100940\relax
\mciteBstWouldAddEndPuncttrue
\mciteSetBstMidEndSepPunct{\mcitedefaultmidpunct}
{\mcitedefaultendpunct}{\mcitedefaultseppunct}\relax
\EndOfBibitem
\bibitem[Stam \latin{et~al.}(2024)Stam, Almeida, Ubbink, van~der Poll, Vogel, Chen, Giordano, Schiettecatte, Hens, and Houtepen]{stam_near-unity_2024}
Stam,~M.; Almeida,~G.; Ubbink,~R.~F.; van~der Poll,~L.~M.; Vogel,~Y.~B.; Chen,~H.; Giordano,~L.; Schiettecatte,~P.; Hens,~Z.; Houtepen,~A.~J. Near-{Unity} {Photoluminescence} {Quantum} {Yield} of {Core}-{Only} {InP} {Quantum} {Dots} via a {Simple} {Postsynthetic} {InF3} {Treatment}. \emph{ACS Nano} \textbf{2024}, \emph{18}, 14685--14695\relax
\mciteBstWouldAddEndPuncttrue
\mciteSetBstMidEndSepPunct{\mcitedefaultmidpunct}
{\mcitedefaultendpunct}{\mcitedefaultseppunct}\relax
\EndOfBibitem
\bibitem[Yuan \latin{et~al.}(2025)Yuan, Tang, Yang, Yu, He, Li, and Li]{yuan_near-infrared-absorbing_2025}
Yuan,~Y.; Tang,~Y.; Yang,~Z.; Yu,~X.; He,~L.; Li,~D.; Li,~W. Near-infrared-absorbing and -emitting indium phosphide quantum dots via nucleation/growth modulation for killing multidrug-resistant bacteria. \emph{Chemical Engineering Journal} \textbf{2025}, \emph{507}, 160729\relax
\mciteBstWouldAddEndPuncttrue
\mciteSetBstMidEndSepPunct{\mcitedefaultmidpunct}
{\mcitedefaultendpunct}{\mcitedefaultseppunct}\relax
\EndOfBibitem
\bibitem[Won \latin{et~al.}(2019)Won, Cho, Kim, Chung, Kim, Chung, Jang, Lee, Kim, and Jang]{won_highly_2019}
Won,~Y.-H.; Cho,~O.; Kim,~T.; Chung,~D.-Y.; Kim,~T.; Chung,~H.; Jang,~H.; Lee,~J.; Kim,~D.; Jang,~E. Highly efficient and stable {InP}/{ZnSe}/{ZnS} quantum dot light-emitting diodes. \emph{Nature} \textbf{2019}, \emph{575}, 634--638\relax
\mciteBstWouldAddEndPuncttrue
\mciteSetBstMidEndSepPunct{\mcitedefaultmidpunct}
{\mcitedefaultendpunct}{\mcitedefaultseppunct}\relax
\EndOfBibitem
\bibitem[Li \latin{et~al.}(2019)Li, Hou, Dai, Yao, Lv, Jin, and Peng]{li_stoichiometry-controlled_2019}
Li,~Y.; Hou,~X.; Dai,~X.; Yao,~Z.; Lv,~L.; Jin,~Y.; Peng,~X. Stoichiometry-{Controlled} {InP}-{Based} {Quantum} {Dots}: {Synthesis}, {Photoluminescence}, and {Electroluminescence}. \emph{Journal of the American Chemical Society} \textbf{2019}, \emph{141}, 6448--6452\relax
\mciteBstWouldAddEndPuncttrue
\mciteSetBstMidEndSepPunct{\mcitedefaultmidpunct}
{\mcitedefaultendpunct}{\mcitedefaultseppunct}\relax
\EndOfBibitem
\bibitem[Guzelian \latin{et~al.}(1996)Guzelian, Katari, Kadavanich, Banin, Hamad, Juban, Alivisatos, Wolters, Arnold, and Heath]{guzelian_synthesis_1996}
Guzelian,~A.~A.; Katari,~J. E.~B.; Kadavanich,~A.~V.; Banin,~U.; Hamad,~K.; Juban,~E.; Alivisatos,~A.~P.; Wolters,~R.~H.; Arnold,~C.~C.; Heath,~J.~R. Synthesis of {Size}-{Selected}, {Surface}-{Passivated} {InP} {Nanocrystals}. \emph{The Journal of Physical Chemistry} \textbf{1996}, \emph{100}, 7212--7219\relax
\mciteBstWouldAddEndPuncttrue
\mciteSetBstMidEndSepPunct{\mcitedefaultmidpunct}
{\mcitedefaultendpunct}{\mcitedefaultseppunct}\relax
\EndOfBibitem
\bibitem[Fu and Zunger(1997)Fu, and Zunger]{fu_inp_1997}
Fu,~H.; Zunger,~A. {InP} quantum dots: {Electronic} structure, surface effects, and the redshifted emission. \emph{Physical Review B} \textbf{1997}, \emph{56}, 1496--1508\relax
\mciteBstWouldAddEndPuncttrue
\mciteSetBstMidEndSepPunct{\mcitedefaultmidpunct}
{\mcitedefaultendpunct}{\mcitedefaultseppunct}\relax
\EndOfBibitem
\bibitem[Cao \latin{et~al.}(2018)Cao, Wang, Wang, Wu, Zhao, and Yang]{cao_layer-by-layer_2018}
Cao,~F.; Wang,~S.; Wang,~F.; Wu,~Q.; Zhao,~D.; Yang,~X. A {Layer}-by-{Layer} {Growth} {Strategy} for {Large}-{Size} {InP}/{ZnSe}/{ZnS} {Core}–{Shell} {Quantum} {Dots} {Enabling} {High}-{Efficiency} {Light}-{Emitting} {Diodes}. \emph{Chemistry of Materials} \textbf{2018}, \emph{30}, 8002--8007\relax
\mciteBstWouldAddEndPuncttrue
\mciteSetBstMidEndSepPunct{\mcitedefaultmidpunct}
{\mcitedefaultendpunct}{\mcitedefaultseppunct}\relax
\EndOfBibitem
\bibitem[Hughes \latin{et~al.}(2019)Hughes, Stein, Friedfeld, Cossairt, and Gamelin]{hughes_effects_2019}
Hughes,~K.~E.; Stein,~J.~L.; Friedfeld,~M.~R.; Cossairt,~B.~M.; Gamelin,~D.~R. Effects of {Surface} {Chemistry} on the {Photophysics} of {Colloidal} {InP} {Nanocrystals}. \emph{ACS Nano} \textbf{2019}, \emph{13}, 14198--14207\relax
\mciteBstWouldAddEndPuncttrue
\mciteSetBstMidEndSepPunct{\mcitedefaultmidpunct}
{\mcitedefaultendpunct}{\mcitedefaultseppunct}\relax
\EndOfBibitem
\bibitem[Sung \latin{et~al.}(2021)Sung, Kim, Yun, Lim, Ko, Jung, Won, Park, Jeon, Lee, Kim, Jun, Sul, and Hwang]{sung_increasing_2021}
Sung,~Y.~M.; Kim,~T.-G.; Yun,~D.-J.; Lim,~M.; Ko,~D.-S.; Jung,~C.; Won,~N.; Park,~S.; Jeon,~W.~S.; Lee,~H.~S.; Kim,~J.-H.; Jun,~S.; Sul,~S.; Hwang,~S. Increasing the {Energy} {Gap} between {Band}-{Edge} and {Trap} {States} {Slows} {Down} {Picosecond} {Carrier} {Trapping} in {Highly} {Luminescent} {InP}/{ZnSe}/{ZnS} {Quantum} {Dots}. \emph{Small} \textbf{2021}, \emph{17}, 2102792\relax
\mciteBstWouldAddEndPuncttrue
\mciteSetBstMidEndSepPunct{\mcitedefaultmidpunct}
{\mcitedefaultendpunct}{\mcitedefaultseppunct}\relax
\EndOfBibitem
\bibitem[Richter \latin{et~al.}(2019)Richter, Binder, Bohn, Grumbach, Neyshtadt, Urban, and Feldmann]{richter_fast_2019}
Richter,~A.~F.; Binder,~M.; Bohn,~B.~J.; Grumbach,~N.; Neyshtadt,~S.; Urban,~A.~S.; Feldmann,~J. Fast {Electron} and {Slow} {Hole} {Relaxation} in {InP}-{Based} {Colloidal} {Quantum} {Dots}. \emph{ACS Nano} \textbf{2019}, \emph{13}, 14408--14415\relax
\mciteBstWouldAddEndPuncttrue
\mciteSetBstMidEndSepPunct{\mcitedefaultmidpunct}
{\mcitedefaultendpunct}{\mcitedefaultseppunct}\relax
\EndOfBibitem
\bibitem[Janke \latin{et~al.}(2018)Janke, Williams, She, Zherebetskyy, Hudson, Wang, Gosztola, Schaller, Lee, Sun, Engel, and Talapin]{janke_origin_2018}
Janke,~E.~M.; Williams,~N.~E.; She,~C.; Zherebetskyy,~D.; Hudson,~M.~H.; Wang,~L.; Gosztola,~D.~J.; Schaller,~R.~D.; Lee,~B.; Sun,~C.; Engel,~G.~S.; Talapin,~D.~V. Origin of {Broad} {Emission} {Spectra} in {InP} {Quantum} {Dots}: {Contributions} from {Structural} and {Electronic} {Disorder}. \emph{Journal of the American Chemical Society} \textbf{2018}, \emph{140}, 15791--15803\relax
\mciteBstWouldAddEndPuncttrue
\mciteSetBstMidEndSepPunct{\mcitedefaultmidpunct}
{\mcitedefaultendpunct}{\mcitedefaultseppunct}\relax
\EndOfBibitem
\bibitem[Jo \latin{et~al.}(2021)Jo, Jo, Choi, Lee, Kim, Yoon, Kim, Han, and Yang]{jo_highly_2021}
Jo,~J.-H.; Jo,~D.-Y.; Choi,~S.-W.; Lee,~S.-H.; Kim,~H.-M.; Yoon,~S.-Y.; Kim,~Y.; Han,~J.-N.; Yang,~H. Highly {Bright}, {Narrow} {Emissivity} of {InP} {Quantum} {Dots} {Synthesized} by {Aminophosphine}: {Effects} of {Double} {Shelling} {Scheme} and {Ga} {Treatment}. \emph{Advanced Optical Materials} \textbf{2021}, \emph{9}, 2100427\relax
\mciteBstWouldAddEndPuncttrue
\mciteSetBstMidEndSepPunct{\mcitedefaultmidpunct}
{\mcitedefaultendpunct}{\mcitedefaultseppunct}\relax
\EndOfBibitem
\bibitem[Lee \latin{et~al.}(2022)Lee, Jo, Kim, Jo, Park, Yang, and Kim]{lee_effectual_2022}
Lee,~Y.; Jo,~D.-Y.; Kim,~T.; Jo,~J.-H.; Park,~J.; Yang,~H.; Kim,~D. Effectual {Interface} and {Defect} {Engineering} for {Auger} {Recombination} {Suppression} in {Bright} {InP}/{ZnSeS}/{ZnS} {Quantum} {Dots}. \emph{ACS Applied Materials \& Interfaces} \textbf{2022}, \emph{14}, 12479--12487\relax
\mciteBstWouldAddEndPuncttrue
\mciteSetBstMidEndSepPunct{\mcitedefaultmidpunct}
{\mcitedefaultendpunct}{\mcitedefaultseppunct}\relax
\EndOfBibitem
\bibitem[Kirkwood \latin{et~al.}(2018)Kirkwood, Monchen, Crisp, Grimaldi, Bergstein, du~Fossé, van~der Stam, Infante, and Houtepen]{kirkwood_finding_2018}
Kirkwood,~N.; Monchen,~J. O.~V.; Crisp,~R.~W.; Grimaldi,~G.; Bergstein,~H. A.~C.; du~Fossé,~I.; van~der Stam,~W.; Infante,~I.; Houtepen,~A.~J. Finding and {Fixing} {Traps} in {II}–{VI} and {III}–{V} {Colloidal} {Quantum} {Dots}: {The} {Importance} of {Z}-{Type} {Ligand} {Passivation}. \emph{Journal of the American Chemical Society} \textbf{2018}, \emph{140}, 15712--15723\relax
\mciteBstWouldAddEndPuncttrue
\mciteSetBstMidEndSepPunct{\mcitedefaultmidpunct}
{\mcitedefaultendpunct}{\mcitedefaultseppunct}\relax
\EndOfBibitem
\bibitem[Stein \latin{et~al.}(2016)Stein, Mader, and Cossairt]{stein_luminescent_2016}
Stein,~J.~L.; Mader,~E.~A.; Cossairt,~B.~M. Luminescent {InP} {Quantum} {Dots} with {Tunable} {Emission} by {Post}-{Synthetic} {Modification} with {Lewis} {Acids}. \emph{The Journal of Physical Chemistry Letters} \textbf{2016}, \emph{7}, 1315--1320\relax
\mciteBstWouldAddEndPuncttrue
\mciteSetBstMidEndSepPunct{\mcitedefaultmidpunct}
{\mcitedefaultendpunct}{\mcitedefaultseppunct}\relax
\EndOfBibitem
\bibitem[Cho \latin{et~al.}(2018)Cho, Kim, Choi, Jang, Min, and Jang]{cho_optical_2018}
Cho,~E.; Kim,~T.; Choi,~S.-m.; Jang,~H.; Min,~K.; Jang,~E. Optical {Characteristics} of the {Surface} {Defects} in {InP} {Colloidal} {Quantum} {Dots} for {Highly} {Efficient} {Light}-{Emitting} {Applications}. \emph{ACS Applied Nano Materials} \textbf{2018}, \emph{1}, 7106--7114\relax
\mciteBstWouldAddEndPuncttrue
\mciteSetBstMidEndSepPunct{\mcitedefaultmidpunct}
{\mcitedefaultendpunct}{\mcitedefaultseppunct}\relax
\EndOfBibitem
\bibitem[Kim \latin{et~al.}(2018)Kim, Zherebetskyy, Bekenstein, Oh, Wang, Jang, and Alivisatos]{kim_trap_2018}
Kim,~T.-G.; Zherebetskyy,~D.; Bekenstein,~Y.; Oh,~M.~H.; Wang,~L.-W.; Jang,~E.; Alivisatos,~A.~P. Trap {Passivation} in {Indium}-{Based} {Quantum} {Dots} through {Surface} {Fluorination}: {Mechanism} and {Applications}. \emph{ACS Nano} \textbf{2018}, \emph{12}, 11529--11540\relax
\mciteBstWouldAddEndPuncttrue
\mciteSetBstMidEndSepPunct{\mcitedefaultmidpunct}
{\mcitedefaultendpunct}{\mcitedefaultseppunct}\relax
\EndOfBibitem
\bibitem[Dümbgen \latin{et~al.}(2021)Dümbgen, Zito, Infante, and Hens]{dumbgen_shape_2021}
Dümbgen,~K.~C.; Zito,~J.; Infante,~I.; Hens,~Z. Shape, {Electronic} {Structure}, and {Trap} {States} in {Indium} {Phosphide} {Quantum} {Dots}. \emph{Chemistry of Materials} \textbf{2021}, \emph{33}, 6885--6896\relax
\mciteBstWouldAddEndPuncttrue
\mciteSetBstMidEndSepPunct{\mcitedefaultmidpunct}
{\mcitedefaultendpunct}{\mcitedefaultseppunct}\relax
\EndOfBibitem
\bibitem[Enright \latin{et~al.}(2022)Enright, Jasrasaria, Hanchard, Needell, Phelan, Weinberg, McDowell, Hsiao, Akbari, Kottwitz, Potter, Wong, Zuo, Atwater, Rabani, and Nuzzo]{enright_role_2022}
Enright,~M.~J. \latin{et~al.}  Role of {Atomic} {Structure} on {Exciton} {Dynamics} and {Photoluminescence} in {NIR} {Emissive} {InAs}/{InP}/{ZnSe} {Quantum} {Dots}. \emph{The Journal of Physical Chemistry C} \textbf{2022}, \emph{126}, 7576--7587\relax
\mciteBstWouldAddEndPuncttrue
\mciteSetBstMidEndSepPunct{\mcitedefaultmidpunct}
{\mcitedefaultendpunct}{\mcitedefaultseppunct}\relax
\EndOfBibitem
\bibitem[Schiettecatte \latin{et~al.}(2024)Schiettecatte, Giordano, Cruyssaert, Bonifas, De~Vlamynck, Van~Avermaet, Zhao, Vantomme, Nayral, Delpech, and Hens]{schiettecatte_enhanced_2024}
Schiettecatte,~P.; Giordano,~L.; Cruyssaert,~B.; Bonifas,~G.; De~Vlamynck,~N.; Van~Avermaet,~H.; Zhao,~Q.; Vantomme,~A.; Nayral,~C.; Delpech,~F.; Hens,~Z. Enhanced {Surface} {Passivation} of {InP}/{ZnSe} {Quantum} {Dots} by {Zinc} {Acetate} {Exposure}. \emph{Chemistry of Materials} \textbf{2024}, \emph{36}, 5996--6005\relax
\mciteBstWouldAddEndPuncttrue
\mciteSetBstMidEndSepPunct{\mcitedefaultmidpunct}
{\mcitedefaultendpunct}{\mcitedefaultseppunct}\relax
\EndOfBibitem
\bibitem[Gwak \latin{et~al.}(2024)Gwak, Shin, Yoo, Seo, Kim, Jang, Lee, Park, Kim, Lim, Kim, Kim, Hwang, and Oh]{gwak_highly_2024}
Gwak,~N.; Shin,~S.; Yoo,~H.; Seo,~G.~W.; Kim,~S.; Jang,~H.; Lee,~M.; Park,~T.~H.; Kim,~B.~J.; Lim,~J.; Kim,~S.~Y.; Kim,~S.; Hwang,~G.~W.; Oh,~N. Highly {Luminescent} {Shell}-{Less} {Indium} {Phosphide} {Quantum} {Dots} {Enabled} by {Atomistically} {Tailored} {Surface} {States}. \emph{Advanced Materials} \textbf{2024}, \emph{36}, 2404480\relax
\mciteBstWouldAddEndPuncttrue
\mciteSetBstMidEndSepPunct{\mcitedefaultmidpunct}
{\mcitedefaultendpunct}{\mcitedefaultseppunct}\relax
\EndOfBibitem
\bibitem[Alexander \latin{et~al.}(2024)Alexander, Kick, McIsaac, and Van~Voorhis]{alexander_understanding_2024}
Alexander,~E.; Kick,~M.; McIsaac,~A.~R.; Van~Voorhis,~T. Understanding {Trap} {States} in {InP} and {GaP} {Quantum} {Dots} through {Density} {Functional} {Theory}. \emph{Nano Letters} \textbf{2024}, \emph{24}, 7227--7235\relax
\mciteBstWouldAddEndPuncttrue
\mciteSetBstMidEndSepPunct{\mcitedefaultmidpunct}
{\mcitedefaultendpunct}{\mcitedefaultseppunct}\relax
\EndOfBibitem
\bibitem[Zhu \latin{et~al.}(2023)Zhu, Bahmani~Jalali, Saleh, Di~Stasio, Prato, Polykarpou, Othonos, Christodoulou, Ivanov, Divitini, Infante, De~Trizio, and Manna]{zhu_boosting_2023}
Zhu,~D.; Bahmani~Jalali,~H.; Saleh,~G.; Di~Stasio,~F.; Prato,~M.; Polykarpou,~N.; Othonos,~A.; Christodoulou,~S.; Ivanov,~Y.~P.; Divitini,~G.; Infante,~I.; De~Trizio,~L.; Manna,~L. Boosting the {Photoluminescence} {Efficiency} of {InAs} {Nanocrystals} {Synthesized} with {Aminoarsine} via a {ZnSe} {Thick}-{Shell} {Overgrowth}. \emph{Advanced Materials} \textbf{2023}, \emph{35}, 2303621\relax
\mciteBstWouldAddEndPuncttrue
\mciteSetBstMidEndSepPunct{\mcitedefaultmidpunct}
{\mcitedefaultendpunct}{\mcitedefaultseppunct}\relax
\EndOfBibitem
\bibitem[Stam \latin{et~al.}(2023)Stam, du~Fossé, Infante, and Houtepen]{stam_guilty_2023}
Stam,~M.; du~Fossé,~I.; Infante,~I.; Houtepen,~A.~J. Guilty as {Charged}: {The} {Role} of {Undercoordinated} {Indium} in {Electron}-{Charged} {Indium} {Phosphide} {Quantum} {Dots}. \emph{ACS Nano} \textbf{2023}, \emph{17}, 18576--18583\relax
\mciteBstWouldAddEndPuncttrue
\mciteSetBstMidEndSepPunct{\mcitedefaultmidpunct}
{\mcitedefaultendpunct}{\mcitedefaultseppunct}\relax
\EndOfBibitem
\bibitem[Ubbink \latin{et~al.}(2022)Ubbink, Almeida, Iziyi, du~Fossé, Verkleij, Ganapathy, van Eck, and Houtepen]{ubbink_water-free_2022}
Ubbink,~R.~F.; Almeida,~G.; Iziyi,~H.; du~Fossé,~I.; Verkleij,~R.; Ganapathy,~S.; van Eck,~E. R.~H.; Houtepen,~A.~J. A {Water}-{Free} {In} {Situ} {HF} {Treatment} for {Ultrabright} {InP} {Quantum} {Dots}. \emph{Chemistry of Materials} \textbf{2022}, \emph{34}, 10093--10103\relax
\mciteBstWouldAddEndPuncttrue
\mciteSetBstMidEndSepPunct{\mcitedefaultmidpunct}
{\mcitedefaultendpunct}{\mcitedefaultseppunct}\relax
\EndOfBibitem
\bibitem[Houtepen \latin{et~al.}(2017)Houtepen, Hens, Owen, and Infante]{houtepen_origin_2017}
Houtepen,~A.~J.; Hens,~Z.; Owen,~J.~S.; Infante,~I. On the {Origin} of {Surface} {Traps} in {Colloidal} {II}–{VI} {Semiconductor} {Nanocrystals}. \emph{Chemistry of Materials} \textbf{2017}, \emph{29}, 752--761\relax
\mciteBstWouldAddEndPuncttrue
\mciteSetBstMidEndSepPunct{\mcitedefaultmidpunct}
{\mcitedefaultendpunct}{\mcitedefaultseppunct}\relax
\EndOfBibitem
\bibitem[Goldzak \latin{et~al.}(2021)Goldzak, McIsaac, and Van~Voorhis]{goldzak_colloidal_2021}
Goldzak,~T.; McIsaac,~A.~R.; Van~Voorhis,~T. Colloidal {CdSe} nanocrystals are inherently defective. \emph{Nature Communications} \textbf{2021}, \emph{12}, 890\relax
\mciteBstWouldAddEndPuncttrue
\mciteSetBstMidEndSepPunct{\mcitedefaultmidpunct}
{\mcitedefaultendpunct}{\mcitedefaultseppunct}\relax
\EndOfBibitem
\bibitem[McIsaac \latin{et~al.}(2023)McIsaac, Goldzak, and Van~Voorhis]{mcisaac_it_2023}
McIsaac,~A.~R.; Goldzak,~T.; Van~Voorhis,~T. It {Is} a {Trap}!: {The} {Effect} of {Self}-{Healing} of {Surface} {Defects} on the {Excited} {States} of {CdSe} {Nanocrystals}. \emph{The Journal of Physical Chemistry Letters} \textbf{2023}, \emph{14}, 1174--1181\relax
\mciteBstWouldAddEndPuncttrue
\mciteSetBstMidEndSepPunct{\mcitedefaultmidpunct}
{\mcitedefaultendpunct}{\mcitedefaultseppunct}\relax
\EndOfBibitem
\bibitem[Bhati \latin{et~al.}(2023)Bhati, Ivanov, Senftle, Tretiak, and Ghosh]{bhati_how_2023}
Bhati,~M.; Ivanov,~S.~A.; Senftle,~T.~P.; Tretiak,~S.; Ghosh,~D. How structural and vibrational features affect optoelectronic properties of non-stoichiometric quantum dots: computational insights. \emph{Nanoscale} \textbf{2023}, \emph{15}, 7176--7185\relax
\mciteBstWouldAddEndPuncttrue
\mciteSetBstMidEndSepPunct{\mcitedefaultmidpunct}
{\mcitedefaultendpunct}{\mcitedefaultseppunct}\relax
\EndOfBibitem
\bibitem[Elward and Chakraborty(2013)Elward, and Chakraborty]{elward_effect_2013}
Elward,~J.~M.; Chakraborty,~A. Effect of {Dot} {Size} on {Exciton} {Binding} {Energy} and {Electron}–{Hole} {Recombination} {Probability} in {CdSe} {Quantum} {Dots}. \emph{Journal of Chemical Theory and Computation} \textbf{2013}, \emph{9}, 4351--4359\relax
\mciteBstWouldAddEndPuncttrue
\mciteSetBstMidEndSepPunct{\mcitedefaultmidpunct}
{\mcitedefaultendpunct}{\mcitedefaultseppunct}\relax
\EndOfBibitem
\bibitem[Kilina \latin{et~al.}(2012)Kilina, Velizhanin, Ivanov, Prezhdo, and Tretiak]{kilina_surface_2012}
Kilina,~S.; Velizhanin,~K.~A.; Ivanov,~S.; Prezhdo,~O.~V.; Tretiak,~S. Surface {Ligands} {Increase} {Photoexcitation} {Relaxation} {Rates} in {CdSe} {Quantum} {Dots}. \emph{ACS Nano} \textbf{2012}, \emph{6}, 6515--6524\relax
\mciteBstWouldAddEndPuncttrue
\mciteSetBstMidEndSepPunct{\mcitedefaultmidpunct}
{\mcitedefaultendpunct}{\mcitedefaultseppunct}\relax
\EndOfBibitem
\bibitem[Xia \latin{et~al.}(2020)Xia, Chen, Zhang, Liu, Wang, Yang, Tang, Lian, He, Liu, Liang, Tan, Gao, Liu, Song, Zhang, Gao, Wang, Lan, Zhang, Müller-Buschbaum, Tang, and Zhang]{xia_facet_2020}
Xia,~Y. \latin{et~al.}  Facet {Control} for {Trap}-{State} {Suppression} in {Colloidal} {Quantum} {Dot} {Solids}. \emph{Advanced Functional Materials} \textbf{2020}, \emph{30}, 2000594\relax
\mciteBstWouldAddEndPuncttrue
\mciteSetBstMidEndSepPunct{\mcitedefaultmidpunct}
{\mcitedefaultendpunct}{\mcitedefaultseppunct}\relax
\EndOfBibitem
\bibitem[Bender \latin{et~al.}(2018)Bender, Raulerson, Li, Goldzak, Xia, Van~Voorhis, Tang, and Roberts]{bender_surface_2018}
Bender,~J.~A.; Raulerson,~E.~K.; Li,~X.; Goldzak,~T.; Xia,~P.; Van~Voorhis,~T.; Tang,~M.~L.; Roberts,~S.~T. Surface {States} {Mediate} {Triplet} {Energy} {Transfer} in {Nanocrystal}–{Acene} {Composite} {Systems}. \emph{Journal of the American Chemical Society} \textbf{2018}, \emph{140}, 7543--7553\relax
\mciteBstWouldAddEndPuncttrue
\mciteSetBstMidEndSepPunct{\mcitedefaultmidpunct}
{\mcitedefaultendpunct}{\mcitedefaultseppunct}\relax
\EndOfBibitem
\bibitem[Zherebetskyy \latin{et~al.}(2015)Zherebetskyy, Zhang, Salmeron, and Wang]{zherebetskyy_tolerance_2015}
Zherebetskyy,~D.; Zhang,~Y.; Salmeron,~M.; Wang,~L.-W. Tolerance of {Intrinsic} {Defects} in {PbS} {Quantum} {Dots}. \emph{The Journal of Physical Chemistry Letters} \textbf{2015}, \emph{6}, 4711--4716\relax
\mciteBstWouldAddEndPuncttrue
\mciteSetBstMidEndSepPunct{\mcitedefaultmidpunct}
{\mcitedefaultendpunct}{\mcitedefaultseppunct}\relax
\EndOfBibitem
\bibitem[Vörös \latin{et~al.}(2017)Vörös, Brawand, and Galli]{voros_hydrogen_2017}
Vörös,~M.; Brawand,~N.~P.; Galli,~G. Hydrogen {Treatment} as a {Detergent} of {Electronic} {Trap} {States} in {Lead} {Chalcogenide} {Nanoparticles}. \emph{Chemistry of Materials} \textbf{2017}, \emph{29}, 2485--2493\relax
\mciteBstWouldAddEndPuncttrue
\mciteSetBstMidEndSepPunct{\mcitedefaultmidpunct}
{\mcitedefaultendpunct}{\mcitedefaultseppunct}\relax
\EndOfBibitem
\bibitem[Nenon \latin{et~al.}(2018)Nenon, Pressler, Kang, Koscher, Olshansky, Osowiecki, Koc, Wang, and Alivisatos]{nenon_design_2018}
Nenon,~D.~P.; Pressler,~K.; Kang,~J.; Koscher,~B.~A.; Olshansky,~J.~H.; Osowiecki,~W.~T.; Koc,~M.~A.; Wang,~L.-W.; Alivisatos,~A.~P. Design {Principles} for {Trap}-{Free} {CsPbX3} {Nanocrystals}: {Enumerating} and {Eliminating} {Surface} {Halide} {Vacancies} with {Softer} {Lewis} {Bases}. \emph{Journal of the American Chemical Society} \textbf{2018}, \emph{140}, 17760--17772\relax
\mciteBstWouldAddEndPuncttrue
\mciteSetBstMidEndSepPunct{\mcitedefaultmidpunct}
{\mcitedefaultendpunct}{\mcitedefaultseppunct}\relax
\EndOfBibitem
\bibitem[Du~Fossé \latin{et~al.}(2022)Du~Fossé, Mulder, Almeida, Spruit, Infante, Grozema, and Houtepen]{du_fosse_limits_2022}
Du~Fossé,~I.; Mulder,~J.~T.; Almeida,~G.; Spruit,~A. G.~M.; Infante,~I.; Grozema,~F.~C.; Houtepen,~A.~J. Limits of {Defect} {Tolerance} in {Perovskite} {Nanocrystals}: {Effect} of {Local} {Electrostatic} {Potential} on {Trap} {States}. \emph{Journal of the American Chemical Society} \textbf{2022}, \emph{144}, 11059--11063\relax
\mciteBstWouldAddEndPuncttrue
\mciteSetBstMidEndSepPunct{\mcitedefaultmidpunct}
{\mcitedefaultendpunct}{\mcitedefaultseppunct}\relax
\EndOfBibitem
\bibitem[Mishra and Ganguli(2016)Mishra, and Ganguli]{mishra_electronic_2016}
Mishra,~S.; Ganguli,~B. Electronic and optical properties of defect \textit{{CdIn}}2\textit{{Te}}4 chalcopyrite semiconductor: {A} first principle approach. \emph{Materials Chemistry and Physics} \textbf{2016}, \emph{173}, 429--437\relax
\mciteBstWouldAddEndPuncttrue
\mciteSetBstMidEndSepPunct{\mcitedefaultmidpunct}
{\mcitedefaultendpunct}{\mcitedefaultseppunct}\relax
\EndOfBibitem
\bibitem[Mishra and Ganguli(2012)Mishra, and Ganguli]{mishra_effect_2012}
Mishra,~S.; Ganguli,~B. Effect of structural distortion and nature of bonding on the electronic properties of defect and {Li}-substituted \textit{{CuInSe}}2 chalcopyrite semiconductors. \emph{Journal of Alloys and Compounds} \textbf{2012}, \emph{512}, 17--22\relax
\mciteBstWouldAddEndPuncttrue
\mciteSetBstMidEndSepPunct{\mcitedefaultmidpunct}
{\mcitedefaultendpunct}{\mcitedefaultseppunct}\relax
\EndOfBibitem
\bibitem[Baimuratov \latin{et~al.}(2017)Baimuratov, Pereziabova, Zhu, Leonov, Baranov, Fedorov, and Rukhlenko]{baimuratov_optical_2017}
Baimuratov,~A.~S.; Pereziabova,~T.~P.; Zhu,~W.; Leonov,~M.~Y.; Baranov,~A.~V.; Fedorov,~A.~V.; Rukhlenko,~I.~D. Optical {Anisotropy} of {Topologically} {Distorted} {Semiconductor} {Nanocrystals}. \emph{Nano Letters} \textbf{2017}, \emph{17}, 5514--5520, Publisher: American Chemical Society\relax
\mciteBstWouldAddEndPuncttrue
\mciteSetBstMidEndSepPunct{\mcitedefaultmidpunct}
{\mcitedefaultendpunct}{\mcitedefaultseppunct}\relax
\EndOfBibitem
\bibitem[Oksenberg \latin{et~al.}(2020)Oksenberg, Merdasa, Houben, Kaplan-Ashiri, Rothman, Scheblykin, Unger, and Joselevich]{oksenberg_large_2020}
Oksenberg,~E.; Merdasa,~A.; Houben,~L.; Kaplan-Ashiri,~I.; Rothman,~A.; Scheblykin,~I.~G.; Unger,~E.~L.; Joselevich,~E. Large lattice distortions and size-dependent bandgap modulation in epitaxial halide perovskite nanowires. \emph{Nature Communications} \textbf{2020}, \emph{11}, 489\relax
\mciteBstWouldAddEndPuncttrue
\mciteSetBstMidEndSepPunct{\mcitedefaultmidpunct}
{\mcitedefaultendpunct}{\mcitedefaultseppunct}\relax
\EndOfBibitem
\bibitem[Cao \latin{et~al.}(2015)Cao, Cheng, Bi, Zhao, Yuan, Liu, Li, Wang, and Che]{cao_crystal_2015}
Cao,~Q.; Cheng,~Y.-F.; Bi,~H.; Zhao,~X.; Yuan,~K.; Liu,~Q.; Li,~Q.; Wang,~M.; Che,~R. Crystal defect-mediated band-gap engineering: a new strategy for tuning the optical properties of {Ag2Se} quantum dots toward enhanced hydrogen evolution performance. \emph{Journal of Materials Chemistry A} \textbf{2015}, \emph{3}, 20051--20055\relax
\mciteBstWouldAddEndPuncttrue
\mciteSetBstMidEndSepPunct{\mcitedefaultmidpunct}
{\mcitedefaultendpunct}{\mcitedefaultseppunct}\relax
\EndOfBibitem
\bibitem[Bertolotti \latin{et~al.}(2016)Bertolotti, Dirin, Ibáñez, Krumeich, Cervellino, Frison, Voznyy, Sargent, Kovalenko, Guagliardi, and Masciocchi]{bertolotti_crystal_2016}
Bertolotti,~F.; Dirin,~D.~N.; Ibáñez,~M.; Krumeich,~F.; Cervellino,~A.; Frison,~R.; Voznyy,~O.; Sargent,~E.~H.; Kovalenko,~M.~V.; Guagliardi,~A.; Masciocchi,~N. Crystal symmetry breaking and vacancies in colloidal lead chalcogenide quantum dots. \emph{Nature Materials} \textbf{2016}, \emph{15}, 987--994\relax
\mciteBstWouldAddEndPuncttrue
\mciteSetBstMidEndSepPunct{\mcitedefaultmidpunct}
{\mcitedefaultendpunct}{\mcitedefaultseppunct}\relax
\EndOfBibitem
\bibitem[Isarov \latin{et~al.}(2017)Isarov, Tan, Bodnarchuk, Kovalenko, Rappe, and Lifshitz]{isarov_rashba_2017}
Isarov,~M.; Tan,~L.~Z.; Bodnarchuk,~M.~I.; Kovalenko,~M.~V.; Rappe,~A.~M.; Lifshitz,~E. Rashba {Effect} in a {Single} {Colloidal} {CsPbBr3} {Perovskite} {Nanocrystal} {Detected} by {Magneto}-{Optical} {Measurements}. \emph{Nano Letters} \textbf{2017}, \emph{17}, 5020--5026\relax
\mciteBstWouldAddEndPuncttrue
\mciteSetBstMidEndSepPunct{\mcitedefaultmidpunct}
{\mcitedefaultendpunct}{\mcitedefaultseppunct}\relax
\EndOfBibitem
\bibitem[Dufour \latin{et~al.}(2019)Dufour, Qu, Greboval, Méthivier, Lhuillier, and Ithurria]{dufour_halide_2019}
Dufour,~M.; Qu,~J.; Greboval,~C.; Méthivier,~C.; Lhuillier,~E.; Ithurria,~S. Halide {Ligands} {To} {Release} {Strain} in {Cadmium} {Chalcogenide} {Nanoplatelets} and {Achieve} {High} {Brightness}. \emph{ACS Nano} \textbf{2019}, \emph{13}, 5326--5334\relax
\mciteBstWouldAddEndPuncttrue
\mciteSetBstMidEndSepPunct{\mcitedefaultmidpunct}
{\mcitedefaultendpunct}{\mcitedefaultseppunct}\relax
\EndOfBibitem
\bibitem[Moscheni \latin{et~al.}(2018)Moscheni, Bertolotti, Piveteau, Protesescu, Dirin, Kovalenko, Cervellino, Pedersen, Masciocchi, and Guagliardi]{moscheni_size-dependent_2018}
Moscheni,~D.; Bertolotti,~F.; Piveteau,~L.; Protesescu,~L.; Dirin,~D.~N.; Kovalenko,~M.~V.; Cervellino,~A.; Pedersen,~J.~S.; Masciocchi,~N.; Guagliardi,~A. Size-{Dependent} {Fault}-{Driven} {Relaxation} and {Faceting} in {Zincblende} {CdSe} {Colloidal} {Quantum} {Dots}. \emph{ACS Nano} \textbf{2018}, \emph{12}, 12558--12570\relax
\mciteBstWouldAddEndPuncttrue
\mciteSetBstMidEndSepPunct{\mcitedefaultmidpunct}
{\mcitedefaultendpunct}{\mcitedefaultseppunct}\relax
\EndOfBibitem
\bibitem[Dolabella \latin{et~al.}(2022)Dolabella, Borzì, Dommann, and Neels]{dolabella_lattice_2022}
Dolabella,~S.; Borzì,~A.; Dommann,~A.; Neels,~A. Lattice {Strain} and {Defects} {Analysis} in {Nanostructured} {Semiconductor} {Materials} and {Devices} by {High}-{Resolution} {X}-{Ray} {Diffraction}: {Theoretical} and {Practical} {Aspects}. \emph{Small Methods} \textbf{2022}, \emph{6}, 2100932\relax
\mciteBstWouldAddEndPuncttrue
\mciteSetBstMidEndSepPunct{\mcitedefaultmidpunct}
{\mcitedefaultendpunct}{\mcitedefaultseppunct}\relax
\EndOfBibitem
\bibitem[Božin \latin{et~al.}(2010)Božin, Malliakas, Souvatzis, Proffen, Spaldin, Kanatzidis, and Billinge]{bozin_entropically_2010}
Božin,~E.~S.; Malliakas,~C.~D.; Souvatzis,~P.; Proffen,~T.; Spaldin,~N.~A.; Kanatzidis,~M.~G.; Billinge,~S. J.~L. Entropically {Stabilized} {Local} {Dipole} {Formation} in {Lead} {Chalcogenides}. \emph{Science} \textbf{2010}, \emph{330}, 1660--1663\relax
\mciteBstWouldAddEndPuncttrue
\mciteSetBstMidEndSepPunct{\mcitedefaultmidpunct}
{\mcitedefaultendpunct}{\mcitedefaultseppunct}\relax
\EndOfBibitem
\bibitem[Guzelturk \latin{et~al.}(2021)Guzelturk, Cotts, Jasrasaria, Philbin, Hanifi, Koscher, Balan, Curling, Zajac, Park, Yazdani, Nyby, Kamysbayev, Fischer, Nett, Shen, Kozina, Lin, Reid, Weathersby, Schaller, Wood, Wang, Dionne, Talapin, Alivisatos, Salleo, Rabani, and Lindenberg]{guzelturk_dynamic_2021}
Guzelturk,~B. \latin{et~al.}  Dynamic lattice distortions driven by surface trapping in semiconductor nanocrystals. \emph{Nature Communications} \textbf{2021}, \emph{12}, 1860\relax
\mciteBstWouldAddEndPuncttrue
\mciteSetBstMidEndSepPunct{\mcitedefaultmidpunct}
{\mcitedefaultendpunct}{\mcitedefaultseppunct}\relax
\EndOfBibitem
\bibitem[Kim \latin{et~al.}(2020)Kim, Heo, Kim, Reboul, Chun, Kang, Bae, Hyun, Lim, Lee, Han, Hyeon, Alivisatos, Ercius, Elmlund, and Park]{kim_critical_2020}
Kim,~B.~H. \latin{et~al.}  Critical differences in {3D} atomic structure of individual ligand-protected nanocrystals in solution. \emph{Science} \textbf{2020}, \emph{368}, 60--67\relax
\mciteBstWouldAddEndPuncttrue
\mciteSetBstMidEndSepPunct{\mcitedefaultmidpunct}
{\mcitedefaultendpunct}{\mcitedefaultseppunct}\relax
\EndOfBibitem
\bibitem[Jana \latin{et~al.}(2017)Jana, de~Frutos, Davidson, and Abécassis]{jana_ligand-induced_2017}
Jana,~S.; de~Frutos,~M.; Davidson,~P.; Abécassis,~B. Ligand-induced twisting of nanoplatelets and their self-assembly into chiral ribbons. \emph{Science Advances} \textbf{2017}, \emph{3}, e1701483\relax
\mciteBstWouldAddEndPuncttrue
\mciteSetBstMidEndSepPunct{\mcitedefaultmidpunct}
{\mcitedefaultendpunct}{\mcitedefaultseppunct}\relax
\EndOfBibitem
\bibitem[Maaten and Hinton(2008)Maaten, and Hinton]{maaten_visualizing_2008}
Maaten,~L. v.~d.; Hinton,~G. Visualizing {Data} using t-{SNE}. \emph{Journal of Machine Learning Research} \textbf{2008}, \emph{9}, 2579--2605\relax
\mciteBstWouldAddEndPuncttrue
\mciteSetBstMidEndSepPunct{\mcitedefaultmidpunct}
{\mcitedefaultendpunct}{\mcitedefaultseppunct}\relax
\EndOfBibitem
\bibitem[Campello \latin{et~al.}(2015)Campello, Moulavi, Zimek, and Sander]{campello_hierarchical_2015}
Campello,~R. J. G.~B.; Moulavi,~D.; Zimek,~A.; Sander,~J. Hierarchical {Density} {Estimates} for {Data} {Clustering}, {Visualization}, and {Outlier} {Detection}. \emph{ACM Trans. Knowl. Discov. Data} \textbf{2015}, \emph{10}, 5:1--5:51\relax
\mciteBstWouldAddEndPuncttrue
\mciteSetBstMidEndSepPunct{\mcitedefaultmidpunct}
{\mcitedefaultendpunct}{\mcitedefaultseppunct}\relax
\EndOfBibitem
\bibitem[Friedman(2001)]{friedman_greedy_2001}
Friedman,~J.~H. Greedy function approximation: {A} gradient boosting machine. \emph{The Annals of Statistics} \textbf{2001}, \emph{29}, 1189--1232\relax
\mciteBstWouldAddEndPuncttrue
\mciteSetBstMidEndSepPunct{\mcitedefaultmidpunct}
{\mcitedefaultendpunct}{\mcitedefaultseppunct}\relax
\EndOfBibitem
\bibitem[Voznyy \latin{et~al.}(2012)Voznyy, Zhitomirsky, Stadler, Ning, Hoogland, and Sargent]{voznyy_charge-orbital_2012}
Voznyy,~O.; Zhitomirsky,~D.; Stadler,~P.; Ning,~Z.; Hoogland,~S.; Sargent,~E.~H. A {Charge}-{Orbital} {Balance} {Picture} of {Doping} in {Colloidal} {Quantum} {Dot} {Solids}. \emph{ACS Nano} \textbf{2012}, \emph{6}, 8448--8455\relax
\mciteBstWouldAddEndPuncttrue
\mciteSetBstMidEndSepPunct{\mcitedefaultmidpunct}
{\mcitedefaultendpunct}{\mcitedefaultseppunct}\relax
\EndOfBibitem
\bibitem[Kühne \latin{et~al.}(2020)Kühne, Iannuzzi, Del~Ben, Rybkin, Seewald, Stein, Laino, Khaliullin, Schütt, Schiffmann, Golze, Wilhelm, Chulkov, Bani-Hashemian, Weber, Borštnik, Taillefumier, Jakobovits, Lazzaro, Pabst, Müller, Schade, Guidon, Andermatt, Holmberg, Schenter, Hehn, Bussy, Belleflamme, Tabacchi, Glöß, Lass, Bethune, Mundy, Plessl, Watkins, VandeVondele, Krack, and Hutter]{kuhne_cp2k_2020}
Kühne,~T.~D. \latin{et~al.}  {CP2K}: {An} electronic structure and molecular dynamics software package - {Quickstep}: {Efficient} and accurate electronic structure calculations. \emph{The Journal of Chemical Physics} \textbf{2020}, \emph{152}, 194103\relax
\mciteBstWouldAddEndPuncttrue
\mciteSetBstMidEndSepPunct{\mcitedefaultmidpunct}
{\mcitedefaultendpunct}{\mcitedefaultseppunct}\relax
\EndOfBibitem
\bibitem[Shao \latin{et~al.}(2015)Shao, Gan, Epifanovsky, Gilbert, Wormit, Kussmann, Lange, Behn, Deng, Feng, Ghosh, Goldey, Horn, Jacobson, Kaliman, Khaliullin, Kuś, Landau, Liu, Proynov, Rhee, Richard, Rohrdanz, Steele, Sundstrom, Woodcock, Zimmerman, Zuev, Albrecht, Alguire, Austin, Beran, Bernard, Berquist, Brandhorst, Bravaya, Brown, Casanova, Chang, Chen, Chien, Closser, Crittenden, Diedenhofen, DiStasio, Do, Dutoi, Edgar, Fatehi, Fusti-Molnar, Ghysels, Golubeva-Zadorozhnaya, Gomes, Hanson-Heine, Harbach, Hauser, Hohenstein, Holden, Jagau, Ji, Kaduk, Khistyaev, Kim, Kim, King, Klunzinger, Kosenkov, Kowalczyk, Krauter, Lao, Laurent, Lawler, Levchenko, Lin, Liu, Livshits, Lochan, Luenser, Manohar, Manzer, Mao, Mardirossian, Marenich, Maurer, Mayhall, Neuscamman, Oana, Olivares-Amaya, O’Neill, Parkhill, Perrine, Peverati, Prociuk, Rehn, Rosta, Russ, Sharada, Sharma, Small, Sodt, Stein, Stück, Su, Thom, Tsuchimochi, Vanovschi, Vogt, Vydrov, Wang, Watson, Wenzel, White, Williams, Yang, Yeganeh, Yost,
  You, Zhang, Zhang, Zhao, Brooks, Chan, Chipman, Cramer, Goddard, Gordon, Hehre, Klamt, Schaefer, Schmidt, Sherrill, Truhlar, Warshel, Xu, Aspuru-Guzik, Baer, Bell, Besley, Chai, Dreuw, Dunietz, Furlani, Gwaltney, Hsu, Jung, Kong, Lambrecht, Liang, Ochsenfeld, Rassolov, Slipchenko, Subotnik, Van~Voorhis, Herbert, Krylov, Gill, and Head-Gordon]{shao_advances_2015}
Shao,~Y. \latin{et~al.}  Advances in molecular quantum chemistry contained in the {Q}-{Chem} 4 program package. \emph{Molecular Physics} \textbf{2015}, \emph{113}, 184--215\relax
\mciteBstWouldAddEndPuncttrue
\mciteSetBstMidEndSepPunct{\mcitedefaultmidpunct}
{\mcitedefaultendpunct}{\mcitedefaultseppunct}\relax
\EndOfBibitem
\bibitem[Kurth \latin{et~al.}(1999)Kurth, Perdew, and Blaha]{kurth_molecular_1999}
Kurth,~S.; Perdew,~J.~P.; Blaha,~P. Molecular and solid-state tests of density functional approximations: {LSD}, {GGAs}, and meta-{GGAs}. \emph{International Journal of Quantum Chemistry} \textbf{1999}, \emph{75}, 889--909\relax
\mciteBstWouldAddEndPuncttrue
\mciteSetBstMidEndSepPunct{\mcitedefaultmidpunct}
{\mcitedefaultendpunct}{\mcitedefaultseppunct}\relax
\EndOfBibitem
\bibitem[Azpiroz \latin{et~al.}(2014)Azpiroz, Ugalde, and Infante]{azpiroz_benchmark_2014}
Azpiroz,~J.~M.; Ugalde,~J.~M.; Infante,~I. Benchmark {Assessment} of {Density} {Functional} {Methods} on {Group} {II}–{VI} {MX} ({M} = {Zn}, {Cd}; {X} = {S}, {Se}, {Te}) {Quantum} {Dots}. \emph{Journal of Chemical Theory and Computation} \textbf{2014}, \emph{10}, 76--89\relax
\mciteBstWouldAddEndPuncttrue
\mciteSetBstMidEndSepPunct{\mcitedefaultmidpunct}
{\mcitedefaultendpunct}{\mcitedefaultseppunct}\relax
\EndOfBibitem
\bibitem[Weigend and Ahlrichs(2005)Weigend, and Ahlrichs]{weigend_balanced_2005}
Weigend,~F.; Ahlrichs,~R. Balanced basis sets of split valence, triple zeta valence and quadruple zeta valence quality for {H} to {Rn}: {Design} and assessment of accuracy. \emph{Physical Chemistry Chemical Physics} \textbf{2005}, \emph{7}, 3297--3305\relax
\mciteBstWouldAddEndPuncttrue
\mciteSetBstMidEndSepPunct{\mcitedefaultmidpunct}
{\mcitedefaultendpunct}{\mcitedefaultseppunct}\relax
\EndOfBibitem
\bibitem[Boys(1960)]{boys_construction_1960}
Boys,~S.~F. Construction of {Some} {Molecular} {Orbitals} to {Be} {Approximately} {Invariant} for {Changes} from {One} {Molecule} to {Another}. \emph{Reviews of Modern Physics} \textbf{1960}, \emph{32}, 296--299\relax
\mciteBstWouldAddEndPuncttrue
\mciteSetBstMidEndSepPunct{\mcitedefaultmidpunct}
{\mcitedefaultendpunct}{\mcitedefaultseppunct}\relax
\EndOfBibitem
\bibitem[Lehtola and Jónsson(2013)Lehtola, and Jónsson]{lehtola_unitary_2013}
Lehtola,~S.; Jónsson,~H. Unitary {Optimization} of {Localized} {Molecular} {Orbitals}. \emph{Journal of Chemical Theory and Computation} \textbf{2013}, \emph{9}, 5365--5372\relax
\mciteBstWouldAddEndPuncttrue
\mciteSetBstMidEndSepPunct{\mcitedefaultmidpunct}
{\mcitedefaultendpunct}{\mcitedefaultseppunct}\relax
\EndOfBibitem
\bibitem[Truhlar(2012)]{truhlar_are_2012}
Truhlar,~D.~G. Are {Molecular} {Orbitals} {Delocalized}? \emph{Journal of Chemical Education} \textbf{2012}, \emph{89}, 573--574\relax
\mciteBstWouldAddEndPuncttrue
\mciteSetBstMidEndSepPunct{\mcitedefaultmidpunct}
{\mcitedefaultendpunct}{\mcitedefaultseppunct}\relax
\EndOfBibitem
\bibitem[Pipek and Mezey(1989)Pipek, and Mezey]{pipek_fast_1989}
Pipek,~J.; Mezey,~P.~G. A fast intrinsic localization procedure applicable for ab initio and semiempirical linear combination of atomic orbital wave functions. \emph{The Journal of Chemical Physics} \textbf{1989}, \emph{90}, 4916--4926\relax
\mciteBstWouldAddEndPuncttrue
\mciteSetBstMidEndSepPunct{\mcitedefaultmidpunct}
{\mcitedefaultendpunct}{\mcitedefaultseppunct}\relax
\EndOfBibitem
\bibitem[Löwdin(1950)]{lowdin_nonorthogonality_1950}
Löwdin,~P. On the {Non}‐{Orthogonality} {Problem} {Connected} with the {Use} of {Atomic} {Wave} {Functions} in the {Theory} of {Molecules} and {Crystals}. \emph{The Journal of Chemical Physics} \textbf{1950}, \emph{18}, 365--375\relax
\mciteBstWouldAddEndPuncttrue
\mciteSetBstMidEndSepPunct{\mcitedefaultmidpunct}
{\mcitedefaultendpunct}{\mcitedefaultseppunct}\relax
\EndOfBibitem
\bibitem[Zhu \latin{et~al.}(2019)Zhu, Thompson, and Martínez]{zhu_geodesic_2019}
Zhu,~X.; Thompson,~K.~C.; Martínez,~T.~J. Geodesic interpolation for reaction pathways. \emph{The Journal of Chemical Physics} \textbf{2019}, \emph{150}, 164103\relax
\mciteBstWouldAddEndPuncttrue
\mciteSetBstMidEndSepPunct{\mcitedefaultmidpunct}
{\mcitedefaultendpunct}{\mcitedefaultseppunct}\relax
\EndOfBibitem
\bibitem[Pedregosa \latin{et~al.}(2011)Pedregosa, Varoquaux, Gramfort, Michel, Thirion, Grisel, Blondel, Prettenhofer, Weiss, Dubourg, Vanderplas, Passos, Cournapeau, Brucher, Perrot, and Duchesnay]{pedregosa_scikit-learn_2011}
Pedregosa,~F. \latin{et~al.}  Scikit-learn: {Machine} {Learning} in {Python}. \emph{Journal of Machine Learning Research} \textbf{2011}, \emph{12}, 2825--2830\relax
\mciteBstWouldAddEndPuncttrue
\mciteSetBstMidEndSepPunct{\mcitedefaultmidpunct}
{\mcitedefaultendpunct}{\mcitedefaultseppunct}\relax
\EndOfBibitem
\bibitem[McInnes \latin{et~al.}(2020)McInnes, Healy, and Melville]{mcinnes_umap_2020}
McInnes,~L.; Healy,~J.; Melville,~J. {UMAP}: {Uniform} {Manifold} {Approximation} and {Projection} for {Dimension} {Reduction}. 2020; \url{http://arxiv.org/abs/1802.03426}\relax
\mciteBstWouldAddEndPuncttrue
\mciteSetBstMidEndSepPunct{\mcitedefaultmidpunct}
{\mcitedefaultendpunct}{\mcitedefaultseppunct}\relax
\EndOfBibitem
\bibitem[Platt \latin{et~al.}(1999)Platt, \latin{et~al.} others]{platt1999probabilistic}
Platt,~J.; others Probabilistic outputs for support vector machines and comparisons to regularized likelihood methods. \emph{Advances in large margin classifiers} \textbf{1999}, \emph{10}, 61--74\relax
\mciteBstWouldAddEndPuncttrue
\mciteSetBstMidEndSepPunct{\mcitedefaultmidpunct}
{\mcitedefaultendpunct}{\mcitedefaultseppunct}\relax
\EndOfBibitem
\bibitem[Breiman(2001)]{breiman_random_2001}
Breiman,~L. Random {Forests}. \emph{Machine Learning} \textbf{2001}, \emph{45}, 5--32\relax
\mciteBstWouldAddEndPuncttrue
\mciteSetBstMidEndSepPunct{\mcitedefaultmidpunct}
{\mcitedefaultendpunct}{\mcitedefaultseppunct}\relax
\EndOfBibitem
\bibitem[Hinton(1990)]{hinton_connectionist_1990}
Hinton,~G.~E. In \emph{Machine {Learning}}; Kodratoff,~Y., Michalski,~R.~S., Eds.; Morgan Kaufmann: San Francisco (CA), 1990; pp 555--610\relax
\mciteBstWouldAddEndPuncttrue
\mciteSetBstMidEndSepPunct{\mcitedefaultmidpunct}
{\mcitedefaultendpunct}{\mcitedefaultseppunct}\relax
\EndOfBibitem
\bibitem[Chawla \latin{et~al.}(2002)Chawla, Bowyer, Hall, and Kegelmeyer]{chawla_smote_2002}
Chawla,~N.~V.; Bowyer,~K.~W.; Hall,~L.~O.; Kegelmeyer,~W.~P. {SMOTE}: {Synthetic} {Minority} {Over}-sampling {Technique}. \emph{Journal of Artificial Intelligence Research} \textbf{2002}, \emph{16}, 321--357\relax
\mciteBstWouldAddEndPuncttrue
\mciteSetBstMidEndSepPunct{\mcitedefaultmidpunct}
{\mcitedefaultendpunct}{\mcitedefaultseppunct}\relax
\EndOfBibitem
\bibitem[He \latin{et~al.}(2008)He, Bai, Garcia, and Li]{he_adasyn_2008}
He,~H.; Bai,~Y.; Garcia,~E.~A.; Li,~S. {ADASYN}: {Adaptive} synthetic sampling approach for imbalanced learning. 2008 {IEEE} {International} {Joint} {Conference} on {Neural} {Networks} ({IEEE} {World} {Congress} on {Computational} {Intelligence}). 2008; pp 1322--1328\relax
\mciteBstWouldAddEndPuncttrue
\mciteSetBstMidEndSepPunct{\mcitedefaultmidpunct}
{\mcitedefaultendpunct}{\mcitedefaultseppunct}\relax
\EndOfBibitem
\bibitem[Breiman(1999)]{breiman_pasting_1999}
Breiman,~L. Pasting {Small} {Votes} for {Classification} in {Large} {Databases} and {On}-{Line}. \emph{Machine Learning} \textbf{1999}, \emph{36}, 85--103\relax
\mciteBstWouldAddEndPuncttrue
\mciteSetBstMidEndSepPunct{\mcitedefaultmidpunct}
{\mcitedefaultendpunct}{\mcitedefaultseppunct}\relax
\EndOfBibitem
\bibitem[Van~Vleck(1932)]{PhysRev.41.208}
Van~Vleck,~J.~H. Theory of the Variations in Paramagnetic Anisotropy Among Different Salts of the Iron Group. \emph{Phys. Rev.} \textbf{1932}, \emph{41}, 208--215\relax
\mciteBstWouldAddEndPuncttrue
\mciteSetBstMidEndSepPunct{\mcitedefaultmidpunct}
{\mcitedefaultendpunct}{\mcitedefaultseppunct}\relax
\EndOfBibitem
\bibitem[Alexander(2025)]{ea_repo}
Alexander,~E. QD ML. \url{https://github.com/troyvvgroup/qd_ml}, 2025\relax
\mciteBstWouldAddEndPuncttrue
\mciteSetBstMidEndSepPunct{\mcitedefaultmidpunct}
{\mcitedefaultendpunct}{\mcitedefaultseppunct}\relax
\EndOfBibitem
\bibitem[Kick \latin{et~al.}(2024)Kick, Alexander, Beiersdorfer, and Van~Voorhis]{kick_super-resolution_2024}
Kick,~M.; Alexander,~E.; Beiersdorfer,~A.; Van~Voorhis,~T. Super-resolution techniques to simulate electronic spectra of large molecular systems. \emph{Nature Communications} \textbf{2024}, \emph{15}, 8001\relax
\mciteBstWouldAddEndPuncttrue
\mciteSetBstMidEndSepPunct{\mcitedefaultmidpunct}
{\mcitedefaultendpunct}{\mcitedefaultseppunct}\relax
\EndOfBibitem
\end{mcitethebibliography}

\end{document}